\documentclass{WileyMSP-template}
\usepackage{cite}

\begin{document}

\pagestyle{fancy}
\rhead{\includegraphics[width=2.5cm]{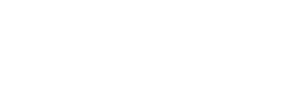}}

\title{Capturing the dynamics of Ti diffusion across Ti$_x$W$_{1-x}$/Cu \newline heterostructures using X-ray photoelectron spectroscopy}

\maketitle


\author{C. Kalha*}
\author{P.~K.~Thakur}%
\author{T.-L.~Lee}%
\author{M.~Reisinger}
\author{J.~Zechner}
\author{M.~Nelhiebel}
\author{A.~Regoutz}


\begin{affiliations}
Curran Kalha, Anna Regoutz\\
Department of Chemistry, University College London, 20 Gordon Street, London, WC1H~0AJ, United Kingdom.\\
Email Address: curran.kalha.19@ucl.ac.uk, a.regoutz@ucl.ac.uk

Pardeep K. Thakur, Tien-Lin Lee\\
Diamond Light Source Ltd., Diamond House, Harwell Science and Innovation Campus, Didcot, OX11 0DE, United Kingdom.

Michael Reisinger, Johannes Zechner, Michael Nelhiebel\\
Kompetenzzentrum Automobil- und Industrie-Elektronik GmbH, Europastraße 8, 9524 Villach, Austria.

\end{affiliations}


\begin{abstract}

Interdiffusion phenomena between adjacent materials are highly prevalent in semiconductor device architectures and can present a major reliability challenge for the industry. To fully capture these phenomena, experimental approaches must go beyond static and post-mortem studies to include in-situ and in-operando setups. Here, soft and hard X-ray photoelectron spectroscopy (SXPS and HAXPES) is used to monitor diffusion in real-time across a proxy device. The device consists of a Si/SiO\textsubscript{2}/Ti$_x$W$_{1-x}$(300~nm)/Cu(25~nm) thin film material stack, with the Ti$_x$W$_{1-x}$ film acting as a diffusion barrier between Si and Cu. The diffusion is monitored through the continuous collection of spectra whilst in-situ annealing to 673~K. Ti within the TiW is found to be highly mobile during annealing, diffusing out of the barrier and accumulating at the Cu surface. Increasing the Ti concentration within the Ti$_x$W$_{1-x}$ film increases the quantity of accumulated Ti, and Ti is first detected at the Cu surface at temperatures as low as 550~K. Surprisingly, at low Ti concentrations ($x$~=~0.054), W is also mobile and diffuses alongside Ti. These results provide crucial evidence for the importance of diffusion barrier composition on their efficacy during device application, delivering insights into the mechanisms underlying their effectiveness and limitations. 

\end{abstract}

\section{Introduction}

The binary pseudo-alloy of titanium-tungsten (Ti$_x$W$_{1-x}$, $x\leq0.3$) is a well-established, effective diffusion barrier and adhesion enhancer within silicon-based semiconductor devices.~\cite{NICOLET1978415, Wang_SQ_1993, ROSHANGHIAS2014386} It is designed to prevent the interdiffusion between adjacent metallisations and the underlying dielectric and semiconductor materials. TiW is compatible with various metallisations (Al, Au, Ag, In and Cu) and has remarkable thermal stability at elevated temperatures ($\leq$850$\degree$C).~\cite{Cunningham_1970, Harris_1976, GHATE1978117, NOWICKI1978195, Olowolafe1985, OPAROWSKI1987313, DIRKS1990201, Misawa_1992, OLOWOLAFE199337, Chiou_1995, BHAGAT20061998, Chang_2000, FUGGER20142487, LePriol2014, Souli_2017, Kalha_TiW_Cu_2022} Consequently, TiW diffusion barriers are now being widely implemented in next-generation SiC-based power semiconductor technologies with copper metallisation schemes,~\cite{Baeri_2004, Behrens_SiC_2013, Liu_2014} and more recently within electrodes for GaAs photoconductive semiconductor switches (PCSSs),~\cite{GaAs} and gate metal stacks in GaN-based high electron mobility transistor (HEMT) devices.~\cite{GaN}\par

Diffusion barriers are needed as Cu and Si readily react at relatively low temperatures to form intermetallic copper-silicide compounds at the interface, which seriously hamper the performance and reliability of devices.~\cite{Corn_1988, Harper_1990, Shacham_Diamand_1993, Liu_1993, Sachdeva_2001, Souli_2017} Studies have shown that TiW films are capable of retarding and limiting this interdiffusion and subsequent reaction.~\cite{Wang_SQ_1993, Souli_2017} However, when subjected to a high thermal budget, a depletion of Ti within the TiW grains has been observed, leading to the accumulation of Ti at grain boundaries.~\cite{CHOOKAJORN2014128} The segregated Ti is then able to diffuse out of the barrier and through the metallisation via grain boundary diffusion.~\cite{Olowolafe1985} This depletion of Ti is thought to lead to a greater defect density within the TiW layer, consequently allowing for the potential of Cu and Si to bypass the barrier and react. Fugger~\textit{et al.} cite that this out-diffusion process is an ``essential factor'' in the failure of this barrier,~\cite{FUGGER20142487} and others have also documented the segregation of Ti during high-temperature annealing.~\cite{OLOWOLAFE199337, OLOWOLAFE199337, Baeri_2004, Plappert_2012, CHOOKAJORN2014128, Kalha_TiW_Cu_2022}\par

Given the importance of the TiW barrier to the overall device performance, reliability and its application in future SiC technologies and beyond, this Ti diffusion degradation process must be better understood, including how it impacts the stability of the TiW/Cu structure. The common thread across the vast majority of past experimental studies on TiW and diffusion barriers in general, including the present authors' previous work,~\cite{Kalha_TiWO, Kalha_TiW_Cu_2022} is that ex-situ samples are used to track the evolution of the diffusion process and to determine the temperature at which the barrier fails. Such studies also often focus on one Ti concentration and are therefore unable to address the effect of the titanium concentration of the film on the degradation mechanism.\par

\begin{figure*}[ht!]
\centering
    \includegraphics[keepaspectratio, width=0.85\linewidth]{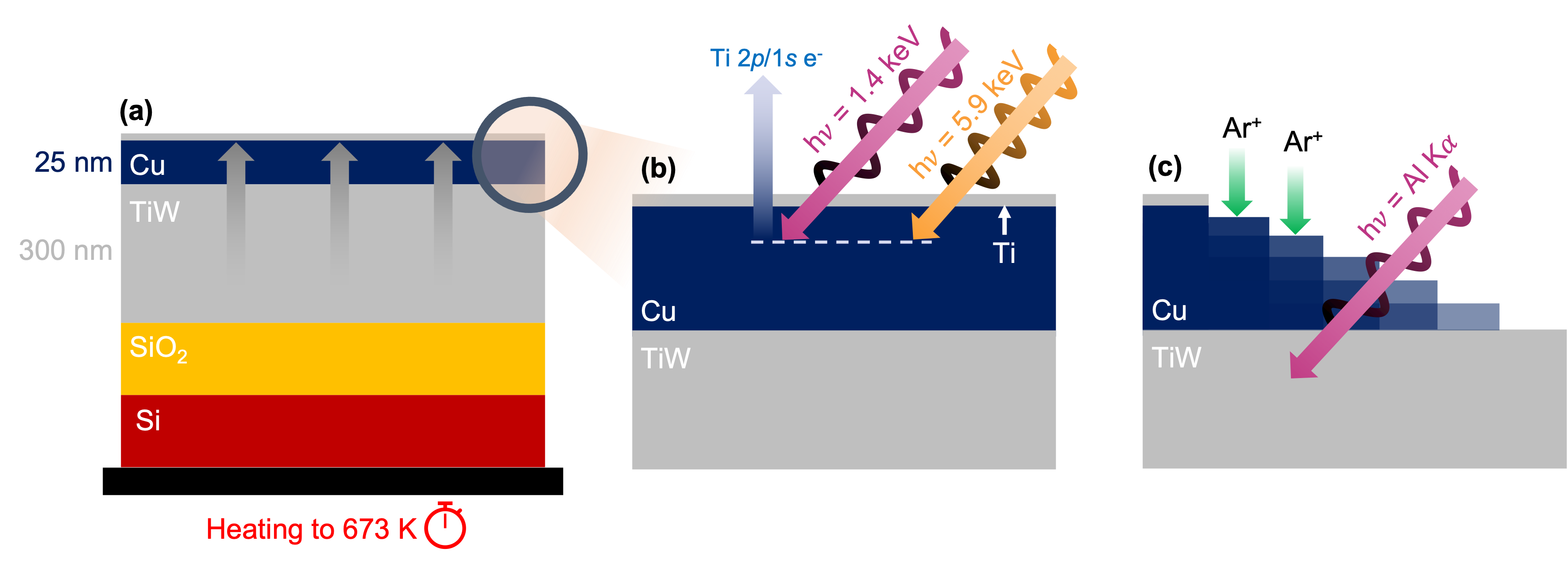}
    \caption{Schematic representation of the samples and experimental approach (not drawn to scale). (a) Device stack on a sample holder being annealed in-situ to 673~K and the expected Ti diffusion represented by grey vertical arrows. (b) A magnified view of the copper surface showing the Ti accumulation and the two photon energies used for SXPS and HAXPES measurements to excite the Ti~2\textit{p} and Ti~1\textit{s} electrons from the same depth. (c) SXPS laboratory-based Ar\textsuperscript{+} sputtering depth profile used to quantify the elemental distribution across the TiW/Cu bilayer after in-situ annealing (i.e. post-mortem).}
    \label{fig:Schematic}
\end{figure*}

Although ex-situ prepared samples give a good representation of the device \emph{after} stress events, it is difficult to correlate the results directly with what a device is experiencing \emph{during} the applied stress.~\cite{Olowolafe1985, Baeri_2004} Therefore, it is crucial to develop new characterisation strategies that can probe the degradation mechanism dynamically under realistic conditions while allowing for changes to the chemical states across the device stack to be monitored.\par

To the best of our knowledge, only Le Priol~\textit{et al.} and Siol~\textit{et al.} provide in-situ monitoring measurements on TiW, both employing in-situ X-ray diffraction (XRD). Le Priol~\textit{et al.} studied the efficiency of a TiW barrier deposited from a 70:30~at.\% W:Ti alloy target against indium diffusion at temperatures between 573-673~K under vacuum.~\cite{LePriol2014} The authors could correlate the TiW barrier efficiency with its microstructure and determine the diffusion coefficient of In in TiW. Siol~\textit{et al.} were interested in understanding the oxidation of TiW alloy precursors, and observed oxygen dissolution and the formation and decomposition of mixed (W,Ti)-oxide phases when ramping the temperature between 303 to 1073~K in air.~\cite{SIOL202095}\par

An explanation for the lack of in-situ/operando experiments in the field, which is in contrast to the importance of these material interfaces in both novel and commercial device applications, is the challenges associated with performing such experiments. These include extensive periods of time required to collect sufficient data, the availability of instruments with in-situ capability, and difficulties in sample preparation and interfacing.\par

The present work combines soft and hard X-ray photoelectron spectroscopies (SXPS and HAXPES) with in-situ annealing to study the effect of annealing temperature, annealing duration, and Ti:W ratio on the thermal stability of TiW/Cu bilayers in real-time, considerably expanding on the existing ex-situ work, including the present authors' previous studies.~\cite{Kalha_TiWO, Kalha_TiW_Cu_2022}  Si/SiO\textsubscript{2}/Ti$_x$W$_{1-x}$(300~nm)/Cu(25~nm) device stacks (see Fig.~\ref{fig:Schematic}(a) for a schematic of the stack) are annealed up to a maximum temperature of 673~K (400$\degree$C) and held there for 5~h. At the same time, soft and hard X-ray photoelectron spectra are continuously recorded to capture the Ti diffusion process and changes to the chemical state across the copper surface (see Fig.~\ref{fig:Schematic}(b) for a schematic). The target temperature of 673~K is selected as it is in a common temperature regime employed during device fabrication to obtain desired grain growth and texture of the copper metallisation.~\cite{Harper_2003, Plappert_2012} Additionally, it is a temperature that can occur at short circuit events during the operation of potential devices.~\cite{NELHIEBEL20111927}\par

A major benefit of combining the two variants of X-ray photoelectron spectroscopy (XPS) is that SXPS is more surface-sensitive, whereas HAXPES enables access to the Ti~1\textit{s} core line. The Ti~1\textit{s} offers an alternative to the commonly measured Ti~2\textit{p} with soft X-ray sources. The Ti~1\textit{s} compared to the Ti~2\textit{p} has the added benefits of covering a smaller binding energy (BE) range and consequently necessitating a shorter collection time, the absence of spin-orbit splitting (SOS), no additional broadening to consider from the Coster-Kronig effect that influences the Ti~2\textit{p}\textsubscript{1/2} peak, and the absence of underlying satellites. For these reasons, the exploitation of the 1\textit{s} core level over the 2\textit{p} is becoming increasingly popular for transition metals, especially for the disentanglement of charge transfer satellite structures in the X-ray photoelectron spectra of metal oxides.~\cite{Woicik_2015, Miedema2015, Ghiasi2019, Woicik_2020, HAXPES_Big_Boy}\par

HAXPES is typically employed as it offers a larger probing depth than conventional SXPS.~\cite{HAXPES_Big_Boy} However, here, it is strategically used to obtain comparable probing depths of the Ti~2\textit{p} and Ti~1\textit{s} core lines, collected with SXPS and HAXPES, respectively. Using this combination, the more widely studied Ti~2\textit{p} spectra can be used to understand the Ti~1\textit{s} spectra better. In addition to the synchrotron-based XPS experiments, quantitative laboratory-based SXPS depth profiles were also conducted on the samples following the in-situ experiment (i.e. post-mortem) to ascertain the quantitative distribution of Ti across the Cu metallisation (see Fig.~\ref{fig:Schematic}(c) for a schematic of the depth profiling).\par

\section{Methodology}

\subsection{Samples}\label{Samples}

Three as-deposited Si/SiO\textsubscript{2}/Ti$_x$W$_{1-x}$/Cu thin film stacks with varying Ti:W composition were prepared through an established industrial route. The stack consists of a 50~nm SiO\textsubscript{2} layer on an un-patterned Si (100) substrate, above which a 300~nm thick TiW layer was deposited via magnetron sputtering. The TiW films were deposited from composite targets with a nominal atomic concentration of 30:70 Ti:W, determined by X-ray fluorescence spectroscopy (XRF). By varying the deposition parameters, three samples with an average Ti concentration, $x$ across the entire film thickness of 5.4$\pm$0.3, 11.5$\pm$0.3, and 14.8$\pm$0.6 at.\% relative to W were realised (e.g (Ti/(Ti+W))$\times$100). These concentrations were determined using laboratory-based SXPS and depth profiling across the entire film thickness (further details regarding the quantification of the TiW films can be found in Supplementary Information I). These samples will be referred to as 5Ti, 10Ti and 15Ti, respectively, for the remainder of the manuscript. Finally, a 25~nm Cu capping layer was deposited via magnetron sputtering on top of the TiW barrier. Deposition of both TiW and Cu was conducted in an argon discharge with no active substrate heating or vacuum break between successive depositions. The deposition chamber operated under a base pressure of 10\textsuperscript{-8}-10\textsuperscript{-7}~mbar. Further details regarding the deposition process have been reported in Refs.~\cite{Plappert_2012, SAGHAEIAN2019137576}.

\subsection{Dynamic synchrotron-based SXPS/HAXPES}

\subsubsection{Beamline optics and end station details}

SXPS and HAXPES measurements were conducted at beamline I09 of the Diamond Light Source, UK,~\cite{Beam2018} at photon energies of 1.415~keV and 5.927~keV, respectively (these will be abbreviated as 1.4~keV and 5.9~keV throughout the remaining manuscript). 1.4~keV was selected using a 400~lines/mm plane grating monochromator, achieving a final energy resolution of 330~meV at room temperature. 5.9~keV was selected using a double-crystal Si~(111) monochromator (DCM) in combination with a post-monochromator Si~(004) channel-cut crystal, achieving a final energy resolution of 290~meV at room temperature. The total energy resolution was determined by extracting the 16/84\% width of the Fermi edge of a clean polycrystalline gold foil (see Supplementary Information II for further information on determining the resolution).~\cite{ISO} The end station of beamline I09 is equipped with an EW4000 Scienta Omicron hemispherical analyser, with a $\pm$28$\degree$ acceptance angle. The base pressure of the analysis chamber was 3.5$\times$10$\textsuperscript{-10}$~mbar. To maximise the efficiency in the collection of spectra, the measurements were conducted in grazing incidence and at near-normal emission geometry.\par 

\subsubsection{Annealing} \label{methods_annealing}

Samples were individually annealed in-situ to a sample target temperature of 673~K (400$\degree$C) using a tungsten filament heater, and held at the temperature for approximately 5~h. The sample plate used for the experiment consisted of a copper disk (3~mm thick, 8~mm diameter) fixed to the centre of a flat tantalum plate, on which the sample was placed and secured using clips. Good thermal contact was made between the copper disk and the sample using a thin silver foil. This allowed the sample temperature to be inferred by attaching an N-type thermocouple to the centre of the copper disc. The thermocouple was also connected to a Lakeshore temperature controller, which was programmed to ramp the sample temperature at a constant rate under a closed-loop control (see Supplementary Information III for an image of the sample plate holder).\par

Prior to in-situ annealing, all samples were gently sputter cleaned in-situ for 10~minutes using a 0.5~keV de-focused argon ion (Ar\textsuperscript{+}) source, operating with a 6~mA emission current and 5$\times$10\textsuperscript{-5}~mbar pressure. This was necessary to remove the native copper oxide that had formed on the sample surface during sample transport.\par

\begin{figure}
\centering
    \includegraphics[keepaspectratio, width=0.4\linewidth]{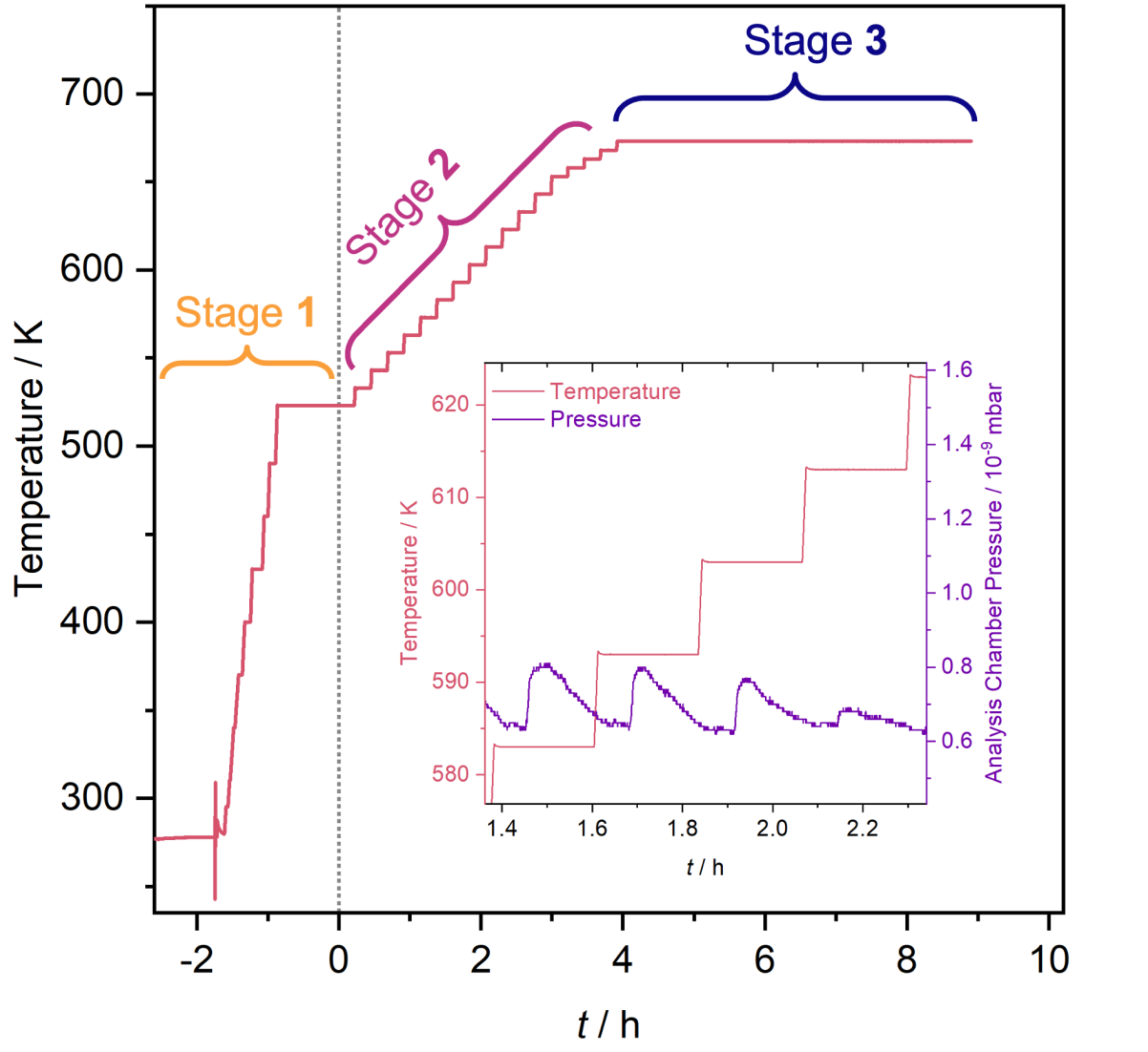}
    \caption{Representative temperature profile acquired from the Lakeshore temperature controller during the measurements on sample 5Ti. The temperature profile consists of three stages. Stage \textbf{1}: a quick ramp to 523~K in a subsidiary chamber. Stage \textbf{2}: a 10~K/[spectral cycle] ramp in the main analysis chamber, which was then decreased to a 5~K/[spectral cycle] ramp once 653~K was reached. The temperature was ramped step-wise in Stage \textbf{2} to allow the pressure in the analysis chamber to recover to $<$7$\times$10\textsuperscript{-10}~mbar after each temperature step (see inset for the pressure profile). Stage \textbf{3}: holding period at 673~K for 5~h. The dotted line at $t$ = 0~h indicates the start of the measurement window.}
    \label{fig:Temp_Profile}
\end{figure}

The process of in-situ annealing encourages the purging of adsorbed gases and organic species within the sample and on the sample surface (i.e. degassing). Therefore, annealing in a UHV environment will increase the chamber pressure, which is undesired, especially during the collection of photoelectron spectra. To account for sample degassing, the annealing process was conducted step-wise to ensure a good analysis chamber pressure was maintained throughout the measurements. Fig.~\ref{fig:Temp_Profile} displays a representative temperature profile acquired for sample 5Ti and the related pressure profile within the analysis chamber (see Supplementary Information IV for the temperature profiles collected for all three samples). The temperature profile consists of three stages. Additionally, as seen in the pressure profile in Fig.~\ref{fig:Temp_Profile}, with every increasing step in temperature, a temporary increase in pressure resulted due to the degassing of the sample.\par

Prior to annealing in the analysis chamber, the samples were first heated in a subsidiary sample preparation chamber to remove the majority of adsorbed molecules. This stage of annealing involves a fast ramp from room temperature to 523~K and will be referred to as Stage \textbf{1} of the annealing process. The Ti diffusion process was assumed to be insignificant in this temperature range. Next, the sample was moved to the main analysis chamber, where the temperature was ramped step-wise from 523 to the target temperature of 673~K while maintaining on average a pressure of 7$\times$10\textsuperscript{-10}~mbar (referred to as Stage \textbf{2}). The temperature was then held at the 673~K target temperature for 5~h (referred to as Stage \textbf{3}). The spectra were continuously collected using SXPS and HAXPES from the start of Stage \textbf{2} until the end of Stage \textbf{3} of the annealing process. The period where the spectra were collected will be referred to as the ``measurement window''. Across the measurement window, the same group of spectra were collected iteratively, which will be referred to as the ``spectral cycle''. Each spectral cycle took approximately 15~minutes to collect, and details on which spectra were selected will be discussed in the following section. During Stage~\textbf{2}, the temperature was increased once a spectral cycle was completed, which coincidentally allows sufficient time for the analysis chamber pressure to recover below 8$\times$10\textsuperscript{-10}~mbar. \par

For completeness, we note that during the initial stages of annealing, sample 10Ti degassed more than samples 5Ti and 15Ti, and therefore the temperature ramp of Stage \textbf{2} for sample 10Ti was paused to allow the pressure to recuperate. This meant that sample 10Ti was held at 543~K for four spectral cycles rather than one. Therefore, the total time of annealing of sample 10Ti was extended by approximately 1~h compared to the annealing time of samples 5Ti and 15Ti. This is not expected to affect the diffusion process significantly or the resultant accumulation profiles, as the Ti diffusion at this temperature is minimal.\par

\subsubsection{Core level selection}\label{Decision}

The spectral cycle, which was run in an iterative loop during the experiment, included the following core level spectra: Cu~2\textit{p}\textsubscript{3/2}, Ti~2\textit{p} and W~4\textit{d} collected with SXPS, and Ti~1\textit{s} collected with HAXPES. The W~4\textit{d} core level was selected over the commonly measured W~4\textit{f} line as the former does not overlap with the core levels of Cu or Ti in this region, whereas the latter overlaps with the Ti~3\textit{p} core level. The Cu Fermi edge was also included in the spectral cycle and was collected with both SXPS and HAXPES throughout the measurement window to (a) provide an intrinsic method of calibrating the BE scale and (b) monitor any change to the total energy resolution as a consequence of raising the sample temperature. Based on 16/84\% fits of the collected Fermi edges across all measurements, the effect of thermal broadening is negligible under the experimental conditions used, and further information can be found in Supplementary Information V. All spectra were aligned to the intrinsic Cu Fermi energy (E\textsubscript{F}) and the spectral areas were obtained using the Thermo Avantage v5.9925 software package. The BE values quoted in this work are considered to have an estimated error of $\pm$0.1~eV.\par

The SXPS photon energy was set to 1.4~keV so that the kinetic energy (KE) of excited Ti~2\textit{p} electrons at this photon energy matches the KE of Ti~1\textit{s} electrons excited with the HAXPES photon energy (KE\textsubscript{Ti~1\textit{s}} $\approx$ KE\textsubscript{Ti~2\textit{p\textsubscript{3/2}}} $\approx$ 961~eV). Using the QUASES software package,~\cite{Shinotsuka_2015} the inelastic mean free path (IMFP) of Ti~2\textit{p} and Ti~1\textit{s} electrons in Cu metal at the SXPS and HAXPES photon energies were calculated. The IMFP for the Ti~1\textit{s} and Ti~2\textit{p}\textsubscript{3/2} is approximately 1.50~nm, and so the estimated probing depth (3$\lambda$) is 4.50~nm. Therefore, a direct comparison between the two Ti core levels will be possible as they originate from very similar probing depths.

\subsection{Laboratory-based SXPS}\label{SXPS_methods}

SXPS depth profile measurements were conducted on the samples that were annealed at I09 using a laboratory-based Thermo K-Alpha+ instrument (i.e. the in-situ annealed samples were removed and kept for a post-mortem analysis). The instrument operates with a monochromated Al~K$\alpha$ photon source ($h\nu$ = 1.4867~keV) and consists of a 180$\degree$ double-focusing hemispherical analyser, a two-dimensional detector that integrates intensity across the entire angular distribution range, and operates at a base pressure of 2$\times$10\textsuperscript{-9}~mbar. A 400~$\mu$m spot size was used for all measurements, achieved using an X-ray anode emission current of 6~mA and a cathode voltage of 12~kV. A flood gun with an emission current of 100~$\mu$A was used to achieve the desired level of charge compensation. The total energy resolution of the spectrometer was determined to be 400~meV. Survey and core level (W~4\textit{f}, Ti~2\textit{p}, O~1\textit{s} and Cu~2\textit{p}\textsubscript{3/2}) spectra were collected with pass energies of 200 and 20~eV, respectively. Depth profiles were conducted using a focused Ar\textsuperscript{+} ion source, operating at 500~eV energy and 10~mA current, rastering over a 2$\times$2~mm\textsuperscript{2} area with a 30$\degree$ sputtering angle. A total of 17 sputter or etch cycles, each lasting 180~s, was carried out with survey and core level spectra collected after each etch cycle. The data were analysed using the Thermo Avantage v5.9925 software package. The error associated with the quantification values is estimated to be $\pm$0.3~at.\% owing to the complexity of the W~4\textit{f} core level and the low quantities of Cu and Ti/W in the TiW and Cu layers, respectively. \par

\section{Results and Discussion}

Reference room temperature survey and core level spectra (Ti~1\textit{s}, Cu~2\textit{p}, Ti~2\textit{p} and W~4\textit{d}) were collected for the three samples after the in-situ sputter cleaning process, and prior to annealing, with the results displayed in Supplementary Information VI. From the survey spectra, the sample surfaces appear clean and are dominated by signals from Cu. Virtually no carbon is detected, and only a trace quantity of oxygen is present when measured with SXPS. The Cu~2\textit{p}\textsubscript{3/2} core level spectra are near identical for the three samples, and the position and line shape are commensurate with metallic copper.~\cite{SCHON197396, SCROCCO197952, Miller_1993} A low-intensity satellite is observed between 943-948~eV in the Cu~2\textit{p}\textsubscript{3/2} core level spectra, but comparing the spectra to reference measurements of a polycrystalline Cu foil and an anhydrous Cu\textsubscript{2}O powder, the satellite intensity is in agreement with the Cu foil. This confirms that the Cu surface of these samples can be considered metallic and the native oxide contribution is minimised after in-situ sputtering.\par

Importantly no Ti or W is observed in these room temperature measurements. This confirms both that the Cu layer is sufficiently thick so that even with SXPS the underlying TiW cannot be probed, and that the surfaces are consistent across all samples. The reference measurements show that the Cu~L\textsubscript{1}M\textsubscript{1}M\textsubscript{4,5} Auger line overlaps with the Ti~1\textit{s} core line but its intensity is vanishingly small.~\cite{COGHLAN1973317, Liu_SpeedyAuger_2021} Nevertheless, care was taken to remove this contribution when we quantified the Ti~1\textit{s} region to accurately determine the relative change in Ti concentration at the surface.\par

The following sections present the Cu, Ti and W core level spectra and associated accumulation profiles as a function of annealing duration/temperature across the three samples, with a focus on the initial stages of annealing and the 673~K holding period.

\subsection{In-situ annealing profiles}


\subsubsection{Copper}\label{Cu}

\begin{figure*}
\centering
    \includegraphics[width=0.9\textwidth,keepaspectratio]{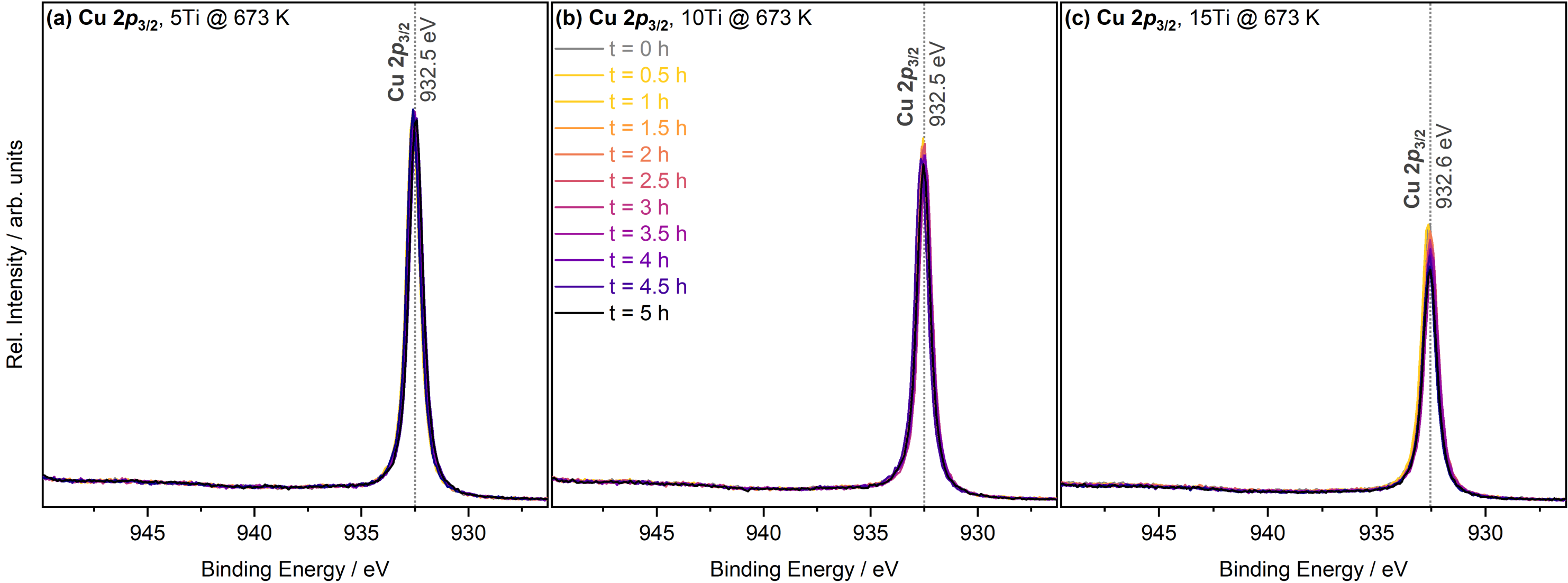}
    \caption{Cu~2\textit{p}\textsubscript{3/2} core level spectra collected during the 673~K holding period (Stage \textbf{3}) for sample (a) 5Ti, (b) 10Ti, and (c) 15Ti. Spectra for each sample are plotted over the same $y$-axis scale to show the differences in intensity across the three samples. The spectra have not been normalised but a constant linear background has been removed. To avoid congestion of this figure, spectra collected every other spectral cycle are presented (i.e. $\approx$30~minutes) rather than at every spectral cycle (i.e. $\approx$15~minutes). The legend displayed in (b) also applies to (a) and (c). Here, $t$~=~0~h refers to the start of the 5~h holding period.}
    \label{fig:CLs_673K_Cu2p}
\end{figure*}

Fig.~\ref{fig:CLs_673K_Cu2p} displays the Cu~2\textit{p}\textsubscript{3/2} core level spectra collected over the 5~h holding period at 673~K for all three samples, i.e. Stage \textbf{3} (with \textit{t}~=~0~h in Fig.~\ref{fig:CLs_673K_Cu2p} referring to the start of the 5~h holding period). The spectra across all samples confirm that Cu still remains in its metallic state during annealing, with a BE position of approximately 932.5~eV. Additionally, the narrow full width at half maximum (FWHM), found to be 0.8~eV, and the lack of significant satellite features in the 943-948~eV region give further confirmation of the metallic nature of the Cu surface.~\cite{SCHON197396, SCROCCO197952, Miller_1993} From Fig.~\ref{fig:CLs_673K_Cu2p} it can be observed that after annealing and within the 673~K holding period, sample 5Ti has the highest Cu~2\textit{p}\textsubscript{3/2} signal intensity (Fig.~\ref{fig:CLs_673K_Cu2p}(a)), followed by samples 10Ti (Fig.~\ref{fig:CLs_673K_Cu2p}(b)) and 15Ti (Fig.~\ref{fig:CLs_673K_Cu2p}(c)). Moreover, within the 5~h holding period, the signal intensity is continually decreasing with annealing duration and this effect is most notable in Fig.~\ref{fig:CLs_673K_Cu2p}(c) for the sample with the highest Ti concentration.\par

\begin{figure*}[ht!]
\centering
    \includegraphics[width=0.95\linewidth,keepaspectratio]{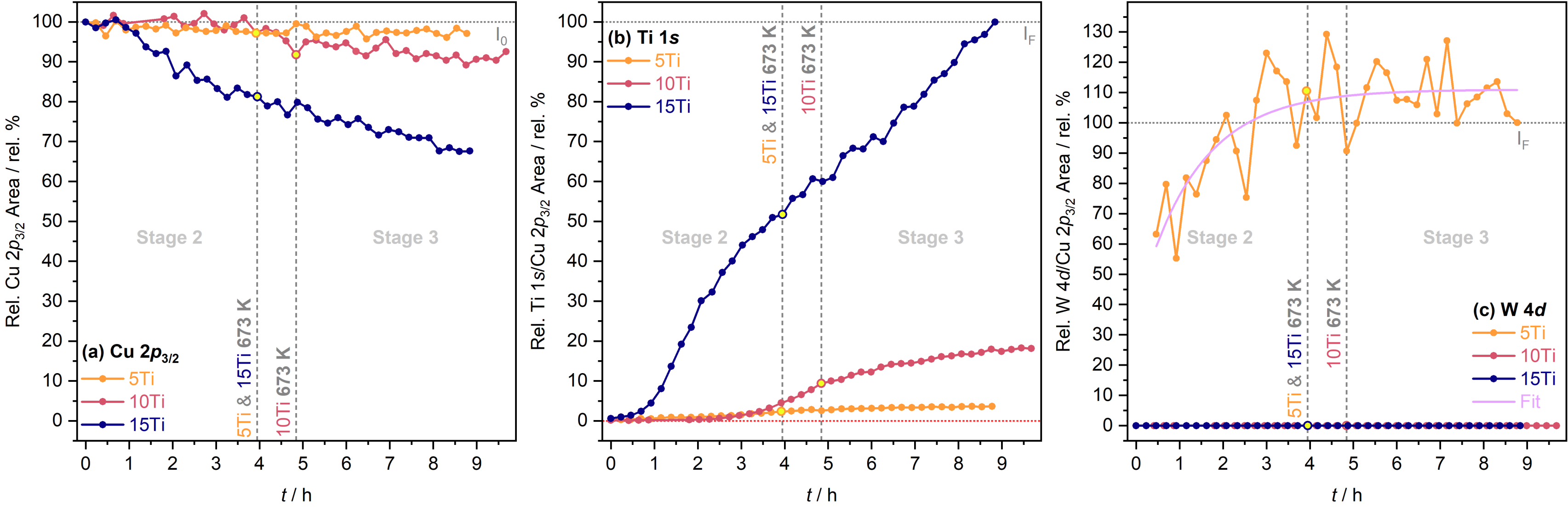}
    \caption{Relative area intensities measured as a function of time, \textit{t} collected across the measurement window for all three samples, including (a) Cu, (b) Ti, and (c) W profiles, determined from peak fitting the Cu~2\textit{p}\textsubscript{3/2}, Ti~1\textit{s} and W~4\textit{d} core level spectra, respectively, at each spectral cycle. Here, $t$~=~0~h refers to the start of the measurement window. The yellow-filled marker for each dataset refers to the time when the 673~K holding period commences (i.e. data points before and after the marker refer to Stage~\textbf{2} and Stage~\textbf{3} of the annealing process). Vertical guidelines are also in place to mark this point for each sample. For Cu, the measured total Cu~2\textit{p}\textsubscript{3/2} areas are normalised relative to the initial raw area (I\textsubscript{0}) of their respective sample (i.e. I/I\textsubscript{0}). For Ti, the measured total raw Ti~1\textit{s} signal area for each sample is first normalised relative to the raw area of the Cu~2\textit{p}\textsubscript{3/2} core level measured during the same spectral cycle and then afterwards the resultant Ti~1\textit{s}/Cu~2\textit{p}\textsubscript{3/2} area is normalised relative to the final raw intensity of sample 15Ti (i.e.~I/I\textsubscript{F}). The W accumulation profile was determined by normalising the measured total raw W~4\textit{d} spectral areas following the method used for the Ti~1\textit{s} normalisation (i.e.~I/I\textsubscript{F}).}
    \label{fig:Composite_Quant}
\end{figure*}

To determine the change in concentration of Cu at the sample surface across the measurement window, peak fit analysis of the Cu~2\textit{p}\textsubscript{3/2} core level was conducted to determine the change in area, with the resultant profile displayed in Fig.~\ref{fig:Composite_Quant}(a). In Fig.~\ref{fig:Composite_Quant}, time, $t$ = 0~h is redefined as the first measurement point of the measurement window (i.e. at the start of Stage \textbf{2} at a temperature of 523~K (250$\degree$C)). Note, $t$ = 0~h in the context of Fig.~\ref{fig:Composite_Quant} is not the same as $t$ = 0~h in Fig.~\ref{fig:CLs_673K_Cu2p}. The same is also true for Fig.~\ref{fig:CLs_673K_Ti1s} and Fig.~\ref{fig:CLs_673K_W4d}, which present the equivalent spectra to Fig.~\ref{fig:CLs_673K_Cu2p} for the Ti~1\textit{s} and W~4\textit{d} core levels, respectively.\par

The Cu~2\textit{p}\textsubscript{3/2} intensity profile in Fig.~\ref{fig:Composite_Quant}(a) reflects what is observed in the core level spectra collected across the 673~K holding period shown in Fig.~\ref{fig:CLs_673K_Cu2p}, in that the Cu~2\textit{p}\textsubscript{3/2} signal intensity decreases as a function of time and annealing temperature across both Stages \textbf{2} and \textbf{3} of the annealing process. The decrease in intensity of the Cu~2\textit{p}\textsubscript{3/2} signal with time is a consequence of the diffusion of Ti out of the TiW layer during annealing. The accumulation of Ti leads to a displacement of Cu atoms and the formation of a Ti-rich surface layer, consequently attenuating the Cu signal. Additionally, when the TiW is more Ti-rich, Fig.~\ref{fig:Composite_Quant}(a) shows that the Cu signal diminishes more extensively suggesting a greater out-diffusion of Ti. As expected based on this interpretation, sample 15Ti shows the largest decay rate in the Cu~2\textit{p}\textsubscript{3/2} signal, followed by sample 10Ti and then 5Ti. At the end of the measurement window, the Cu~2\textit{p}\textsubscript{3/2} signal intensity has depreciated by approximately 2.8, 8.8 and 32.3~\%, for sample 5Ti, 10Ti and 15Ti, respectively. \par

\subsubsection{Titanium} \label{sec:Ti}

The Ti~1\textit{s} core level spectra collected across the 5~h 673~K holding period (Stage \textbf{3}) are displayed in Fig.~\ref{fig:CLs_673K_Ti1s}, with the BE positions of the main signals annotated (see Supplementary Information VII and VIII for the equivalent Ti~2\textit{p} core level spectra and heat maps of the Ti~1\textit{s} spectra collected across the measurement window, respectively). \par

Fig.~\ref{fig:CLs_673K_Ti1s} shows that by the time the 673~K holding period starts, a Ti~1\textit{s} peak is observed across all three samples and the intensity continually increases during the 5~h holding period. This confirms that the onset of diffusion occurs prior to Stage \textbf{3} of the annealing process as assumed during the discussion of the Cu profile. Significant differences in intensity of the Ti~1\textit{s} spectra as a function of Ti concentration are observed, with sample 15Ti showing a considerably more intense peak than sample 10Ti and 5Ti (note the $\times$30 magnification of the 5Ti spectra). Notably, the spectral line shape also appears different across the samples indicating a change in the chemical state of the accumulated Ti. \par

\begin{figure*}
\centering
    \includegraphics[width=0.9\textwidth,keepaspectratio]{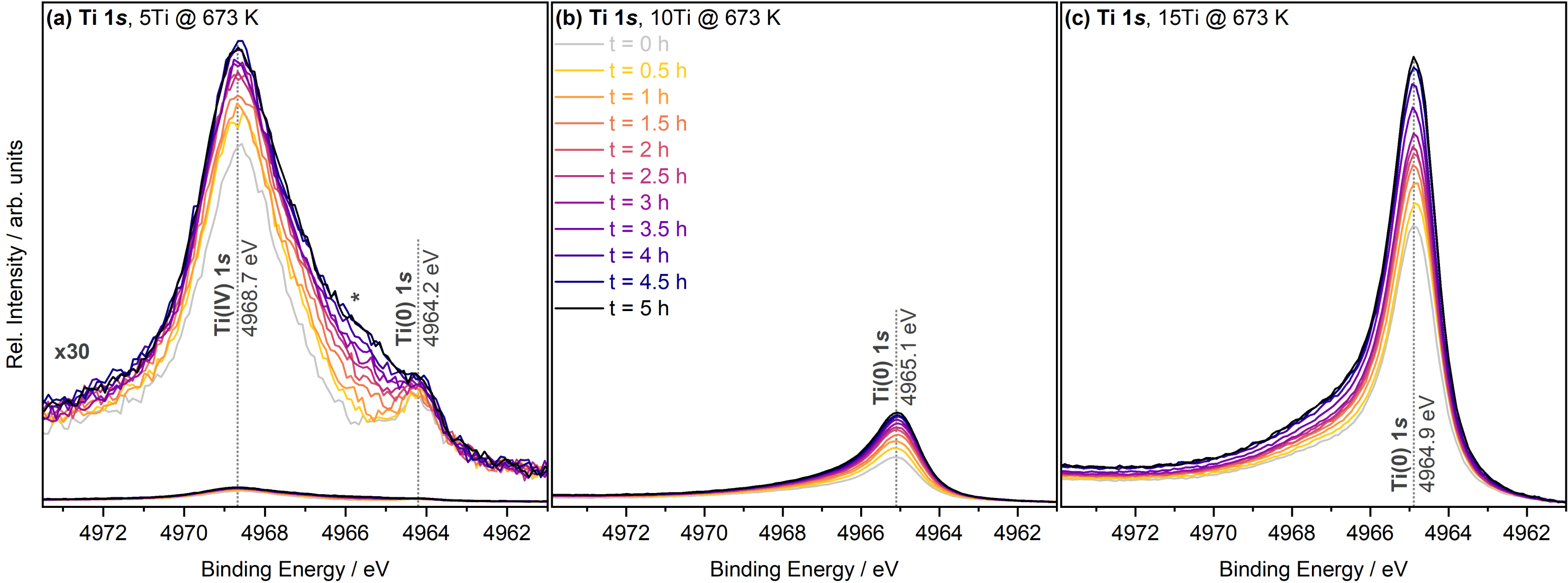}
    \caption{Ti~1\textit{s} core level spectra collected during the 673~K holding period (Stage \textbf{3}) for sample (a) 5Ti, (b) 10Ti, and (c) 15Ti. Spectra for each core level are plotted over the same $y$-axis scale to show the differences in intensity across the three samples. The spectra have not been normalised, but a constant linear background has been removed. Additionally, spectra recorded every other spectral cycle are displayed to aid with the interpretation of the data. For sample 5Ti, the spectra are also shown magnified by $\times$30 to aid with viewing. The legend displayed in (b) also applies to (a) and (c). Here, $t$~=~0~h refers to the start of the 5~h holding period.}
    \label{fig:CLs_673K_Ti1s}
\end{figure*}

All spectra exhibit a lower BE feature at BEs of 4964.2-4965.1~eV, corresponding to metallic Ti in varying environments (labelled as Ti(0)). As the Ti~1\textit{s} core level is not as widely studied as Ti~2\textit{p} due to the need for hard X-ray sources, only a handful of publications exist, with reported BEs varying considerably.~\cite{hagstrom1964extension, Nordberg1966, Diplas_2001, diplas2001electron, MOSLEMZADEH2006129, Woicik_2015, Renault_2018, RISTERUCCI201778, Regoutz_2018} The BE positions of the Ti(0)~1\textit{s} peak observed in the present work fall within the literature range of metallic Ti, and the asymmetric line shape of the peak, which can be clearly observed in Fig.~\ref{fig:CLs_673K_Ti1s}(b) and (c), is commensurate with this assignment. An asymmetric line shape is a hallmark of the core level spectra of many transition metals.~\cite{HUFNER1975417}

The 10Ti and 15Ti samples show a small BE difference of 0.2~eV, which could be attributed to the differences in the Ti:Cu and/or Ti:O ratio at the evolving surface. In contrast, the BE position in the 5Ti spectra is considerably lower, with a -0.9~eV shift relative to the BE position observed in the spectra of sample 10Ti. This shift can be attributed to the distinctly different surface configuration of this sample due to the dominance of Ti-O environments and the co-diffusion of tungsten, both of which will be discussed later. Moreover, the quantity of Ti diffused to the surface is incredibly small for sample 5Ti, and therefore, the shift could be due to strong surface effects, with far fewer nearest neighbours being Ti leading to a negative shift in BE position.~\cite{CHOPRA1986L311, Kuzmin_2011} \par

During the 673~K holding period, the nature of the accumulated Ti for samples 10Ti and 15Ti is predominately metallic, given that a single asymmetric peak is visible (see Figs.~\ref{fig:CLs_673K_Ti1s}(b) and (c)). The accumulated Ti for sample 5Ti, shown in Fig.~\ref{fig:CLs_673K_Ti1s}(a) is strikingly different as the intensity of the lower BE metallic peak is overshadowed by a large, fairly symmetric peak at approximately +4.5~eV from the Ti(0)~1\textit{s} peak. This peak, labelled as Ti(IV)~1\textit{s}, is attributed to Ti-O environments in the Ti 4+ oxidation state (i.e. TiO\textsubscript{2} like). Renault~\textit{et al.} report the Ti~1\textit{s} BE position of the TiO\textsubscript{2} environment on a TiN film at 4968.8~eV,~\cite{Renault_2018} which agrees well with the value reported here. Therefore, unlike samples 10Ti and 15Ti, the Ti accumulated at the surface of sample 5Ti is not predominately metallic but oxidic. Additionally, there is a shoulder on the lower BE side of this Ti(IV)~1\textit{s} peak (marked with an asterisk, *), which is attributed to lower valence states of Ti (i.e. 2+, 3+) that may also form due to the limited quantity of oxygen expected to be present (see Supplementary Information IX for a peak fit analysis of the spectra highlighting the presence of such environments). This shoulder increases in intensity with increasing annealing duration, and at the end of the 5~h period, a distinct Ti(0)~1\textit{s} peak is difficult to observe.\par
 
To aid with the interpretation of the Ti~1\textit{s} spectra, as well as validate the chemical state assignments made so far, the Ti~2\textit{p} spectra are used in parallel (see Supplementary Information VII). The Ti~2\textit{p} spectra for samples 10Ti and 15Ti show a doublet peak with an asymmetric line shape at 454.5 and 460.6~eV (SOS = 6.1~eV), in agreement with  metallic Ti.~\cite{TANAKA1990429, Kuznetsov_1992} For sample 5Ti, three peaks are identified at 453.8, 459.0, and 464.8~eV. The lowest BE peak corresponds to Ti~2\textit{p}\textsubscript{3/2} of Ti(0), whereas the other two correspond to the doublet of Ti oxide in the 4+ oxidation state (SOS = 5.8~eV), labelled as Ti(IV) (with the Ti(IV)~2\textit{p}\textsubscript{3/2} peak overlapping the Ti(0)~2\textit{p}\textsubscript{1/2} peak). These BE positions and the SOS of the Ti(IV) oxide doublet match well with literature values.~\cite{Diebold_1996, Regoutz_2016}\par

A shift of the lower BE Ti(0)~2\textit{p}\textsubscript{3/2} peak between the three samples is observed, with the peak positioned at 453.8, 454.7 and 454.4~eV for sample 5Ti, 10Ti and 15Ti, respectively. The relative shifts are similar to those observed in the Ti~1\textit{s} spectra. Moreover, the Ti~2\textit{p} spectra recorded for sample 5Ti also display a shoulder on the lower BE side of the main Ti(IV)~2\textit{p}\textsubscript{3/2}, again reflecting what has been observed in the Ti~1\textit{s} spectra, suggesting the presence of lower valence oxidation states that may form during the reaction between Ti and oxygen.~\cite{POUILLEAU1997235, MCCAFFERTY199992} Overall, this confirms the peak assignments made using the Ti~1\textit{s} core level are valid and shows the importance of using multiple core levels to have confidence in the assignment of chemical states.\par


The observation of almost completely oxidised Ti on the surface of sample 5Ti is of interest, given that these measurements were conducted under ultra-high vacuum (UHV) conditions and annealed in-situ. The level of observed oxidation cannot be explained by Ti gettering residual oxygen from the analysis chamber as the quantity present in the chamber is insufficient to promote oxidation of Ti to the extent observed. Furthermore, as the sample is heated during the measurement, the sticking coefficients for adsorbed gases are greatly reduced. An alternative source of oxygen is residual oxygen within the Cu film, whether that be intrinsic to the film (i.e. incorporated during deposition) or that the sputtering process prior to annealing did not fully remove the native oxide layer that formed during the exposure of the samples to the atmosphere. From the room temperature reference survey spectra found in Supplementary Information VI, a small intensity O~1\textit{s} signal is present. Laboratory-based SXPS depth profiling on the as-deposited samples was conducted to determine the oxygen level within the starting (i.e. pre-annealed) films and to validate this assumption further. Three sputter cycles (or etch steps) were completed before the underlying TiW signal became strong (see Supplementary Information X for the collected spectra). The profiles showed that within the Cu bulk, less than 2~rel. at.\% of O is present, i.e., $<$2~at.\% O, $>$98~at.\% Cu. Within the errors of the performed quantification, this amount would be enough to facilitate the observed Ti oxidation.\par

\begin{figure*}[ht]
\centering
    \includegraphics[width=0.65\linewidth,keepaspectratio]{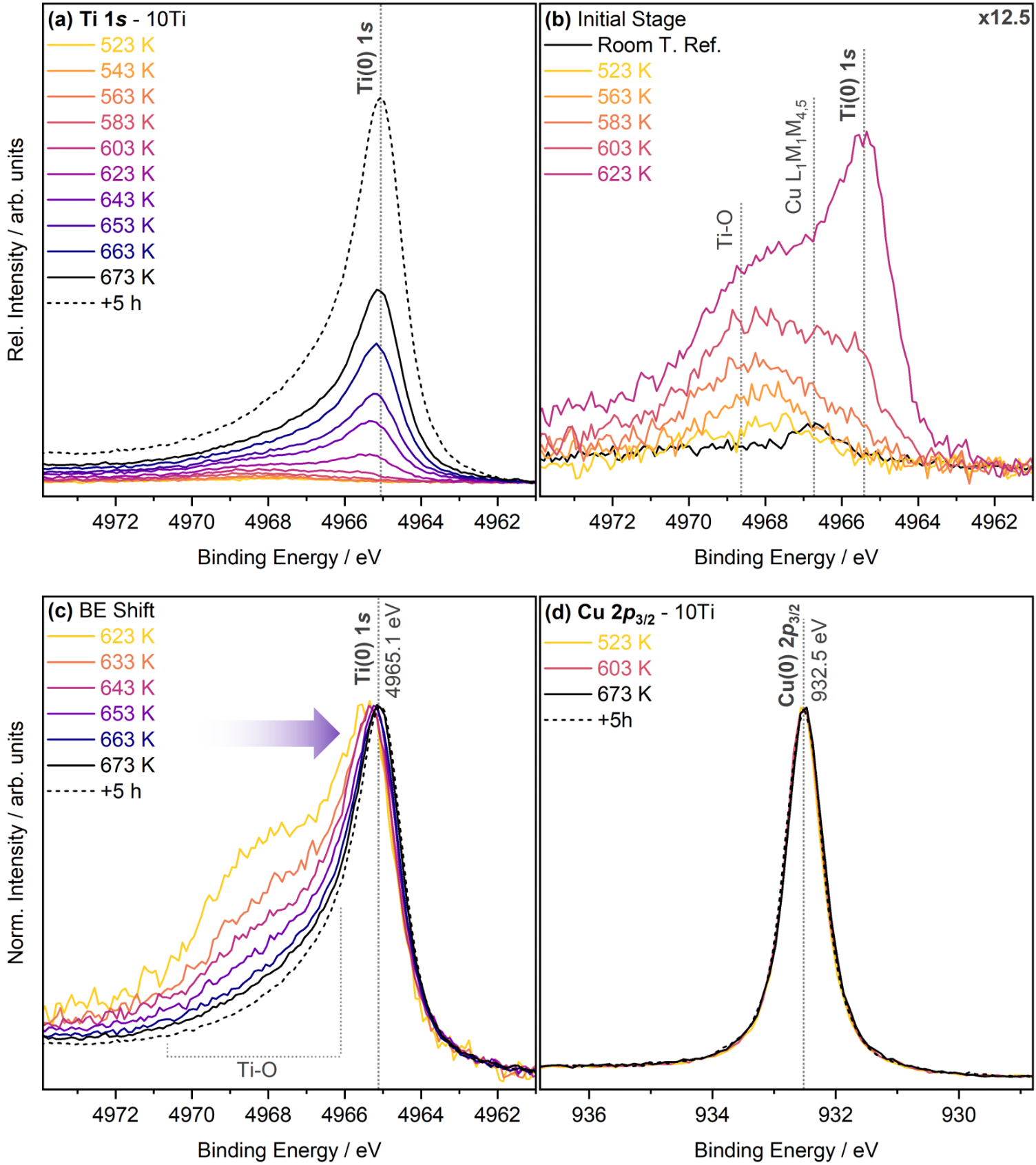}
    \caption{Initial stages of annealing (523-673~K) described by the Cu~2\textit{p}\textsubscript{3/2} and Ti~1\textit{s} core level spectra. (a) Raw Ti~1\textit{s} core level spectra collected (i.e. with no intensity normalisation) at each temperature increment, with +5~h referring to the data collected at the end of the 5~h 673~K holding period. (b) A magnified view of the raw Ti~1\textit{s} core level spectra collected between 523-623~K and a room temperature reference measurement on the same sample (i.e. before annealing) to highlight the Cu Auger contribution. (c) Normalised (0-1) Ti~1\textit{s} core level spectra to emphasise the change in line shape as a function of temperature. (d) Normalised (0-1) Cu~2\textit{p}\textsubscript{3/2} spectra taken at selected temperatures. (a) and (b), and (c) and (d) are plotted on the same $y$-axis scale, respectively (note the $\times$12.5 magnification of the $y$-axis scale of (b)).}
    \label{fig:01}
\end{figure*}

Overall it is apparent that the oxidation of Ti is dependent on both the quantity and rate of accumulation of Ti metal at the surface. Given the significant Ti oxidation observed for sample 5Ti, owing to the low concentration of accumulated Ti, it would be expected that during the early stages of annealing for the higher concentration samples, when an equally low concentration of Ti is expected to accumulate, oxidation should also occur. To confirm this and explore the oxidation of accumulated Ti further, Fig.~\ref{fig:01} displays the Ti~1\textit{s} core level spectra collected across the measurement window for sample 10Ti (equivalent figures for sample 5Ti and 15Ti can be viewed in Supplementary Information XI and XII, respectively). Fig.~\ref{fig:01}(a) shows that during the initial stages of annealing sample 10Ti ($\leq$603~K), the intensity first increases within the region of 4966-4970~eV. After 603~K the intensity increases below 4966~eV, where the metallic Ti(0)~1\textit{s} peak is located, and this peak quickly becomes the dominant contribution to the total line shape and consequently masks the intensity of the environments seen on the higher BE side.\par

From Fig.~\ref{fig:CLs_673K_Ti1s}(a), we know that the 4966-4970~eV region corresponds to Ti-O environments, namely the Ti(IV) oxidation environment, suggesting that even for sample 10Ti, during the initial stages of annealing when the accumulated Ti concentration is low, oxidation of Ti metal occurs. This region will be referred to as Ti-O environments in the following discussion. Fig.~\ref{fig:01}(b) further emphasises the development of Ti-O environments by focusing on the spectra collected between 523-623~K. From this, it is clear that Ti-O environments evolve first and then after 603~K, the Ti(0)~1\textit{s} peak appears due to the continuing diffusion of Ti metal from the TiW layer. It should be noted that the Cu~LMM Auger peak is also present in this region, however, given that the main Cu~2\textit{p}\textsubscript{3/2} core level peak decreases with annealing duration and temperature, the observed increase in spectral intensity in this region cannot be explained by any interference from the Auger peak.\par

The transition from predominantly Ti oxide to metal is evident in Fig.~\ref{fig:01}(c), showing the Ti~1\textit{s} spectra normalised to the maximum peak height. This figure shows that the main intensity peak signal shifts towards lower BEs across the temperature range of 623-673~K (highlighted with an arrow), and this is accompanied by a decrease in the relative intensity of the Ti-O region. The observed shift is due to the emergence of the Ti(0)~1\textit{s} metal peak and the overall reduction of the Ti-O contribution to the total spectral line shape. Lastly, Fig.~\ref{fig:01}(d) displays the Cu~2\textit{p}\textsubscript{3/2} spectrum recorded at different temperatures across the measurement window, and no discernible change is observed in the spectra. Additionally, Supplementary Information XIII shows that the same observation is true when comparing the Cu~2\textit{p}\textsubscript{3/2} line shape across all three samples. This indicates that only the Ti, not the Cu, is undergoing changes to its chemical state at the developing interface.\par

Therefore, oxidation of the surface accumulated Ti is also observed in sample 10Ti but is more evident during the initial stages of annealing where the rate of metal Ti diffusion and quantity of accumulated Ti is small. The same holds true for sample 15Ti as seen in Supplementary Information XII. 
Beyond the qualitative analysis of the Ti~1\textit{s}/2\textit{p} spectra, an accumulation profile of Ti at the Cu surface across the measurement window can be obtained. The Ti accumulation profiles for the samples were extracted from the Ti~1\textit{s} core level spectral areas and are displayed in Fig.~\ref{fig:Composite_Quant}(b) (the equivalent Ti~2\textit{p} profile can be found in Supplementary Information XIV). Before discussing these profiles, it is important to reiterate that they represent changes in the quantity of surface-accumulated Ti with respect to time and not temperature, but with increasing time, the temperature also rises.\par

The temperature at which Ti is first observed at the Cu surface (i.e.~the onset), is difficult to identify with full confidence as the signal is very small, especially for samples 5Ti and 10Ti. For these two samples, the temperature range between 553-563~K (i.e. within the first two hours of Stage \textbf{2}) is when a Ti signal is clearly detectable. The detection of these small Ti signals was only possible through analysing the Ti~1\textit{s} core level as it was much more intense and sharper than the Ti~2\textit{p} (Supplementary Information XV provides a comparison of the Ti~2\textit{p} and Ti~1\textit{s} measured at the same point to highlight this issue). In contrast, for sample 15Ti, it is obvious from Fig.~S15(b) in Supplementary Information XII, that Ti is observed from the start of the measurement window (i.e.~523~K) and may have even begun to accumulate during Stage \textbf{1} of the annealing process.  \par 

The Ti profile displayed in Fig.~\ref{fig:Composite_Quant}(b) shows that with increasing the concentration of Ti within the TiW film, a greater out-diffusion of Ti is observed and thus, a greater accumulation of Ti on the Cu surface occurs. From the profile, it is apparent that the rate of diffusion and the quantity of accumulated Ti differs significantly across the three samples. Focusing on the last data point in the Ti profile at the end of the 673~K holding period, the Ti~1\textit{s}/Cu~2\textit{p}\textsubscript{3/2} area ratios of samples 5Ti and 10Ti are 3.7$\pm$0.5 and 18.2$\pm$0.5~\%, respectively of that of sample 15Ti. This indicates that a linear relationship between the Ti concentration in the film and the quantity of accumulated Ti on the Cu surface does not exist (i.e. they do not scale proportionally).\par

Sinojiya~\textit{et al.} studied similar Ti$_x$W$_{1-x}$ films across a composition range and observed that above a certain Ti concentration threshold, segregation of Ti toward the grain boundaries was favoured, and this enrichment increased with increasing Ti concentration.~\cite{Sinojiya_2022} Additionally, they observed that the change in Ti concentration not only enhances the segregation of Ti but is also accompanied by a change in stress, microstructure, and grain boundary density within the TiW films. A columnar grain boundary structure was also observed at higher concentrations with a relatively higher grain boundary density. Therefore, in our case, for sample 15Ti it is possible that a greater quantity of Ti was already segregated from the TiW grains within the as-deposited films or that annealing promoted a greater segregation compared to samples 5Ti and 10Ti, and consequently that this led to the differences observed in the Ti accumulation profile between the three samples. Furthermore, based on the work of Sinojiya~\textit{et al.}, the expected differences in the microstructure across samples 5, 10 and 15Ti will also contribute to the changes observed in the Ti diffusion profile as properties such as grain boundary density will affect the rate of diffusion. \par

The Ti accumulation profile displayed in Fig.~\ref{fig:Composite_Quant}(b), collected across the measurement window of all three samples, exhibit two different diffusion regimes. The first regime occurs before the 673~K target is reached (i.e. during Stage \textbf{2}), wherein a rapid exponential increase in intensity occurs when ramping the temperature. Once the 673~K target is reached (i.e. during Stage \textbf{3}), the second regime occurs wherein the diffusion rate begins to decelerate and starts to plateau. A plateau is observed for sample 5Ti, and signs of a plateau are present for sample 10Ti by the end of the measurement window. In contrast, the profile for sample 15Ti does not show signs of plateauing, indicating that Ti continues to accumulate at the Cu surface under the temperature and measurement window tested in this experiment. By fitting the linear portions of the Ti~1\textit{s} profile collected during Stages \textbf{2} and \textbf{3} of annealing, the rate of increase in the Ti~1\textit{s} signal intensity relative to sample 15Ti can be determined. The results of the linear fits of Stage \textbf{2} for samples 5Ti, 10Ti and 15Ti were found to be 0.7, 4.9 and 16.5, respectively, and for Stage \textbf{3} were found to be 0.2, 1.4 and 9.7, respectively (error estimated to be $\pm$20\%). These values highlight the dramatic decrease in the Ti accumulation rate during Stage \textbf{3} of annealing. Multiple processes could be responsible for these changes in the accumulation rate. For instance, only a finite quantity of Ti may be available to segregate from the TiW grains, therefore, after annealing for several hours, a plateau is reached as no more Ti is available to diffuse.~\cite{Kalha_TiW_Cu_2022} Additionally, the accumulation appears to decelerate after the 673~K mark is reached. This deceleration may imply that when subjected to a constant temperature rather than a temperature ramp, the rate of diffusion levels off as a steady-state system is reached due to the thermal input remaining at a constant rate. \par


\subsubsection{Tungsten}
%

\begin{figure}
\centering
    \includegraphics[width=0.33\linewidth,keepaspectratio]{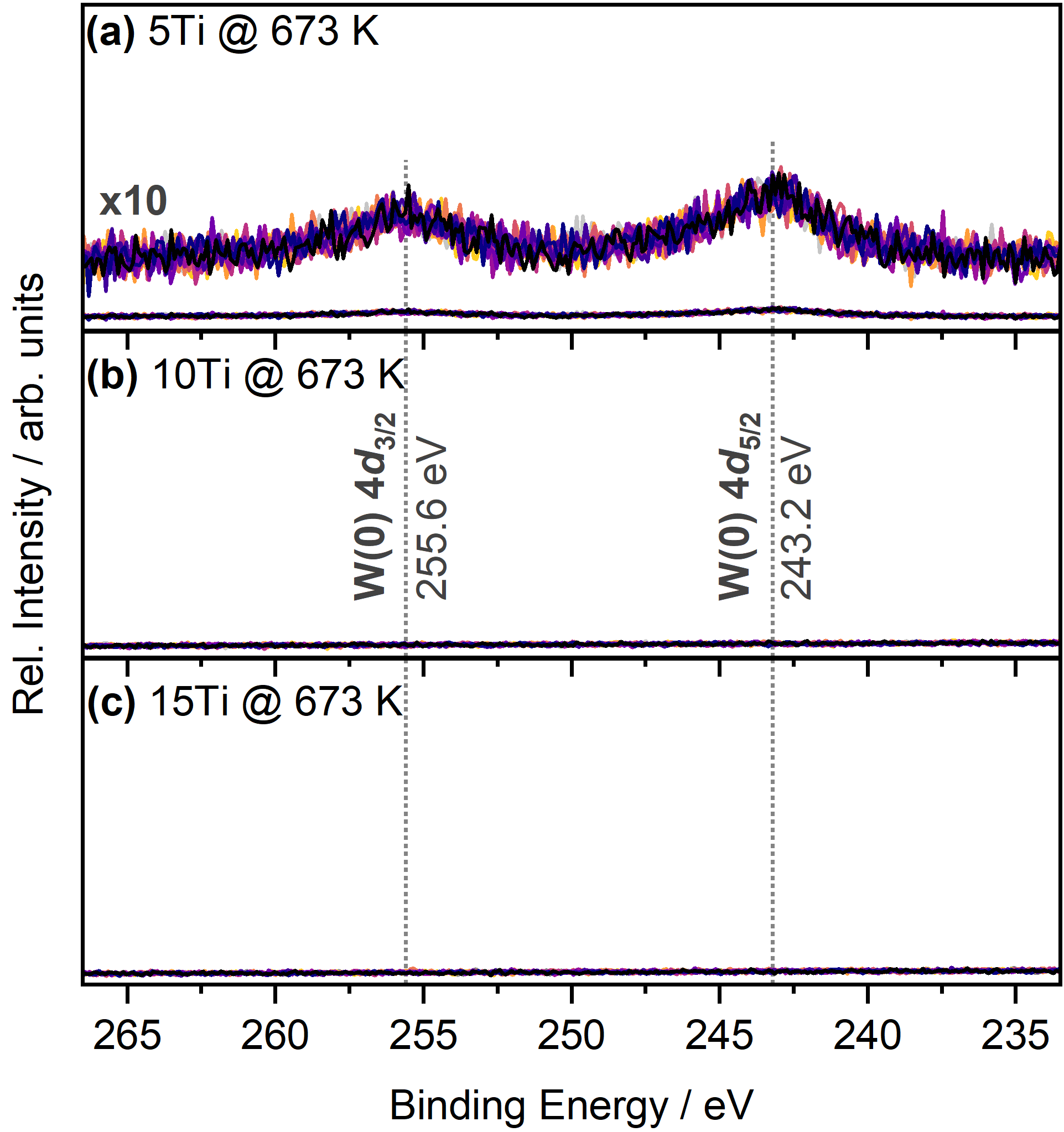}
    \caption{W~4\textit{d} core level spectra collected during the 673~K holding period (Stage \textbf{3}) for samples (a) 5Ti, (b) 10Ti, and (c) 15Ti. Spectra for each core level are plotted over the same $y$-axis scale to show the differences in intensity across the three samples. Note the $\times$10 magnification of the spectra for sample 5Ti in (a). The spectra have not been normalised, but a constant linear background has been removed. Additionally, spectra recorded every other spectral cycle are displayed to aid with the interpretation of the data. For sample 5Ti (a), the inset shows a $\times$10 magnification of the spectra to aid with viewing. The legend is the same as that used in Fig.~\ref{fig:CLs_673K_Cu2p}(b) and Fig.~\ref{fig:CLs_673K_Ti1s}(b). Here, $t$~=~0~h refers to the start of the 5~h holding period.}
    \label{fig:CLs_673K_W4d}
\end{figure}

Fig.~\ref{fig:CLs_673K_W4d} displays the collected W~4\textit{d} core level spectra for all samples during the 5~h 673~K holding period (Stage \textbf{3}). W is not observed within this period for the 10Ti and 15Ti samples, however, it is detected for sample 5Ti, whose TiW film contains the lowest Ti concentration. This confirms that W co-diffuses to the surface only for sample 5Ti, and given that it is already detected at $t$ = 0~h of the holding period, the diffusion likely occurred prior to Stage \textbf{3}. The BE position of the W~4\textit{d}\textsubscript{5/2} peak is at 243.2~eV, in good agreement with metallic W.~\cite{Kalha_W_2022} Within the 5~h period, the concentration of surface accumulated W does not increase in intensity with increasing annealing duration, suggesting that the accumulation has plateaued and the diffusion has subsided. The presence of W at the Cu surface may also influence the oxidation behaviour of the accumulated Ti as observed in the previous section.\par

Fig.~\ref{fig:Composite_Quant}(c) displays the relative accumulation profile of W at the Cu surface across the measurement window for all three samples. Due to the poor signal-to-noise ratio (SNR) of the W~4\textit{d} spectra, it is difficult to have complete confidence in determining the exact temperature at which W is first observed for sample 5Ti. However, the signal becomes apparent at 553-563~K, similar to when Ti was observed at the surface of the same sample. The poor SNR is also responsible for the large scatter in the accumulation profile, leading to an area change greater than 100~rel.\%. Fitting the data points with an asymptotic curve shows that a plateau is reached when crossing from Stage \textbf{2} to Stage \textbf{3} of the annealing process, with the 673~K holding period profile flattening, similar to what was observed for the Ti profile. The observed plateau indicates that a finite quantity of W is able to migrate from the barrier and that a steady state is reached within the measurement window explored. \par 

The diffusion of W is surprising as the vast majority of studies on TiW only report the out-diffusion of Ti. For example, even studies on pure W diffusion barriers,~\cite{Shen_1986, mercier1997, GUPTA1975362, wang_1994} or on a TiW barrier with a relatively low Ti concentration (4.9~at.\%)~\cite{Evans_1994} do not report any mobility of W. However, some studies observe W diffusion from a W or TiW barrier within thin film stacks at temperatures below 600$\degree$C, although no details are given on a possible reason as to why this occurs.~\cite{Christou_1975, Palmstrom_1985, ASHKENAZI1993746} \par

Based on the present results, it is hypothesised that the Ti concentration of the TiW film dictates the overall stability of the diffusion barrier. If it is too low (i.e. in the 5Ti sample), a small amount of W becomes mobile and is free to migrate through the Cu overlayer alongside Ti and accumulate at the surface. This suggests that Ti plays an active role in stabilising the barrier and achieving the desired microstructure necessary for good barrier performance. Therefore, tuning the Ti concentration to an optimum value can significantly improve the barrier performance.

\subsection{Elemental distribution across the in-situ annealed TiW/Cu bilayer} \label{DP}

From the in-situ annealing results, it is clear that under the conditions tested, the out-diffusion of Ti from TiW and through the Cu metallisation is observed for the two samples with the higher Ti concentration - 10Ti and 15Ti. Whereas, for the lowest Ti concentration sample (5Ti), both Ti and W diffuse through the copper metallisation. To quantify the elemental ratio of Cu, Ti, and W across the metallisation, depth profiling using laboratory-based SXPS was conducted on the in-situ annealed samples (i.e. post-mortem). Survey spectra collected at each etch cycle for all three samples can be found in Supplementary Information XVI, showing the change in composition and transition between the Cu overlayer and TiW sublayer. 

\begin{figure*}[ht]
\centering
    \includegraphics[keepaspectratio, width=0.95\linewidth]{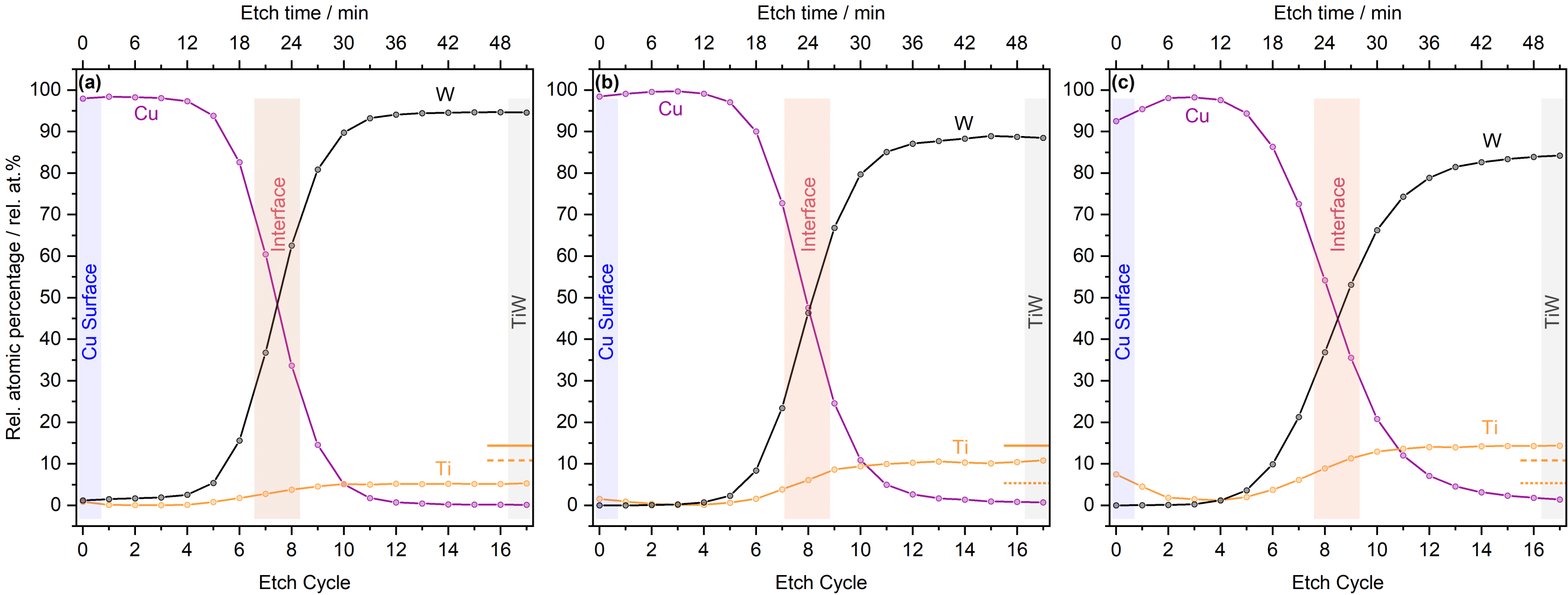}
    \caption{Post-mortem laboratory-based SXPS sputter depth profiles collected across samples (a) 5Ti, (b) 10Ti and (c) 15Ti after in-situ annealing at beamline I09. Etch Cycle 0 refers to the spectra collected on the as-received sample (i.e. before any sputtering). Horizontal guidelines are added to show the final Ti~at.\% for each sample, with the dotted, dashed and solid orange lines referring to samples 5, 10 and 15Ti, respectively.}
    \label{fig:DP}
\end{figure*}

The depth profiles for the three samples displayed in Fig.~\ref{fig:DP} highlight the distribution of Ti across the Cu layer and confirm what was observed in the in-situ measurements, in that at the Cu surface, the quantity of accumulated Ti increases in intensity as the Ti concentration of the film increases. The profiles further confirm that Ti is found throughout the Cu film after annealing. However, its distribution is not uniform, with more Ti observed at the Cu/air and TiW/Cu interfaces. Despite the strong out-diffusion, distinct Cu and TiW zones are still observable in the depth profiles, showing that the TiW/Cu bilayer has not failed when stressed under these conditions.\par

Several studies on Cu/Ti bilayer films have identified that a reaction between the two films can occur as low as 325$\degree$C, leading to the formation of intermetallic CuTi and Cu\textsubscript{3}Ti compounds at the interface.~\cite{Liotard_1985, Li_1992, Apblett_1992} As shown in Fig.~\ref{fig:01}(c), the shifts observed for the Ti~1\textit{s} core line are representative of a changing oxide to metal ratio rather than the formation of an intermetallic compound, whereas the Cu~2\textit{p}\textsubscript{3/2} spectra displayed in Fig.~\ref{fig:01}(d) show no change in the line shape. If an intermetallic compound were to form, one would expect some systematic change to the spectra with increasing annealing duration and temperature or for samples with a higher Ti concentration in the TiW film, as these will cause the greatest surface enrichment of Ti on the Cu. The possibility of such a reaction is difficult to answer from the core level spectra alone. The depth profiles can aid with this discussion. At etch cycle 0 (i.e. as-received surface), the Ti:Cu ratio for sample 15Ti is 7.5:92.5. Of course, this may be slightly skewed as the surface is oxidised, and so there may be additional diffusion of Ti across the metal/oxide interface, but also a carbon surface layer is present which will affect the quantification. Nevertheless, this ratio is insufficient to form stoichiometric CuTi or Cu\textsubscript{3}Ti intermetallic phases that were reported in previous studies on the Ti/Cu interface.~\cite{Liotard_1985} Therefore, based on this literature, the presented spectra and the quantified Ti:Cu ratio, a reaction between Cu and Ti at the developing Cu/Ti interface does not occur due to the relatively small amount of diffused Ti, which again may explain why no systematic shifts in the core level spectra commensurate with a Cu-Ti reaction were observed. However, it should be noted that it may not be possible to observe intermetallic compounds as (a) the quantity of diffused Ti is very small, and (b) the Cu~2\textit{p}\textsubscript{3/2} core line is known to have small chemical shifts.~\cite{Chawla1992DiagnosticSF}\par

In terms of W, the depth profiles shown in Fig.~\ref{fig:DP} confirm that W is only observed at the Cu surface for sample 5Ti and is not present at the surface or within the Cu bulk for samples 10Ti and 15Ti. Fig.~\ref{fig:DP}(a) shows that for sample 5Ti, the W profile is fairly constant across etch cycles 0-3, suggesting that W is homogeneously distributed throughout the Cu metallisation and is not accumulated at the Cu/air interface like Ti. Quantification of the Cu, Ti and W signals reveals that at the surface of sample 5Ti (etch cycle 0), the composition is 97.9 (Cu), 0.9 (Ti), and 1.2 (W)~rel. at.\%, showing that significant W diffusion has occurred. \par

Fig.~\ref{fig:DP} shows that the Cu signal tends towards 0~rel. at.\% for all samples when the interface is reached. However, Cu is still detected at the deepest point of the depth profile, with a composition at etch cycle 17 calculated to be 0.1 (Cu) 99.9 (Ti + W), 0.7 (Cu) 99.3 (Ti + W), and 1.4 (Cu) 98.6 (Ti + W) rel. at.\%, for samples 5Ti, 10Ti and 15Ti, respectively. Moreover, with increasing Ti concentration, the element profiles broaden, and their gradients toward the ``interface'' labelled zone reduce. This provides evidence that there is a degree of intermixing at the TiW/Cu interface, and for films with higher Ti concentrations, a greater intermixing is observed due to the larger rate of atomic flux of Ti across the interface during annealing. Therefore, the out-diffusion of Ti from the TiW also promotes the down diffusion of Cu into the TiW layer, and consequently, the TiW and Cu layers bleed into each other. \par

To summarise, the depth profiles show that clear TiW and Cu zones remain across all samples despite the diffusion and intermixing that occurs during annealing. Although the concentration of Cu observed at the deepest point of the depth profiles increases when the concentration of Ti in the TiW increases, it is difficult to determine how deep the Cu diffuses, as the measurement point of the last depth profile etch cycle is still very much at the surface of the 300~nm thick TiW film. However, given the low concentration of Cu detected at this point ($\leq$1.4~at.\%), and the fact that distinct Cu and TiW zones still remain, one can be confident that under the conditions tested, the TiW barrier has not failed, and the majority of Cu is held above the barrier.\par

\section{Conclusion}

The thermal stability of the TiW barrier in conjunction with a Cu metallisation overlayer was evaluated in real-time using a combination of SXPS and HAXPES, and annealing the sample in-situ to a target temperature of 673~K. The primary mode of degradation was the segregation of Ti from the TiW barrier and its diffusion to the copper surface to form a surface overlayer. The concentration of Ti in TiW was shown to have a significant influence on the thermal stability of the TiW barrier. Two thresholds are observed when moving across the TiW composition window tested here: (I) below a certain concentration of Ti, W gains mobility, suggesting that the incorporation of Ti stabilises W, and (II) above a certain concentration of Ti the diffusion drastically increases, suggesting that at higher concentrations grain boundary segregation of Ti from the TiW grains is favoured, resulting in significantly more out-diffusion of Ti. The post-mortem depth profiles validate the effectiveness of TiW diffusion barriers as despite the degradation observed during annealing, the Ti depletion is not significant enough to lead to the failure of the barrier, as distinct Cu and TiW zones are still present. Overall, it is clear that the composition heavily dictates the stability of TiW, but under the conditions tested, all three barrier compositions remain effective at suppressing the permeation of copper. Based on this, the TiW alloy can cement itself as an excellent diffusion barrier to incorporate into future device technologies.

\medskip
\textbf{Supporting Information} \par 
The Supplementary Information includes room temperature reference spectra, heat maps of the Ti~1\textit{s} spectra collected across the measurement window, and the Ti~2\textit{p} spectra collected for all samples during the 673~K holding period. Additionally, core level spectra collected for samples 5Ti and 15Ti during the 523-673~K annealing period, survey spectra from the laboratory-based SXPS depth profile, information on the residual level of oxygen within the Cu films from laboratory-based SXPS, and a comparison of the Ti~2\textit{p} and Ti~1\textit{s} core levels  can be found in the Supplementary Information. Information on the peak fitting procedures used, and the method to determine and monitor the thermal broadening is also available in the Supplementary Information.

\medskip
\textbf{Acknowledgements} \par 
C.K. acknowledges the support from the Department of Chemistry, UCL. A.R. acknowledges the support from the Analytical Chemistry Trust Fund for her CAMS-UK Fellowship. This work was carried out with the support of Diamond Light Source, instrument I09 (proposal NT29451-1 and NT29451-2). The authors would like to thank Dave McCue, I09 beamline technician, for his support of the experiments.
\medskip

\bibliographystyle{MSP}
\bibliography{references}

\clearpage

\begin{figure}
\centering
\textbf{Table of Contents}\\
\medskip
  \includegraphics[height= 7cm,keepaspectratio]{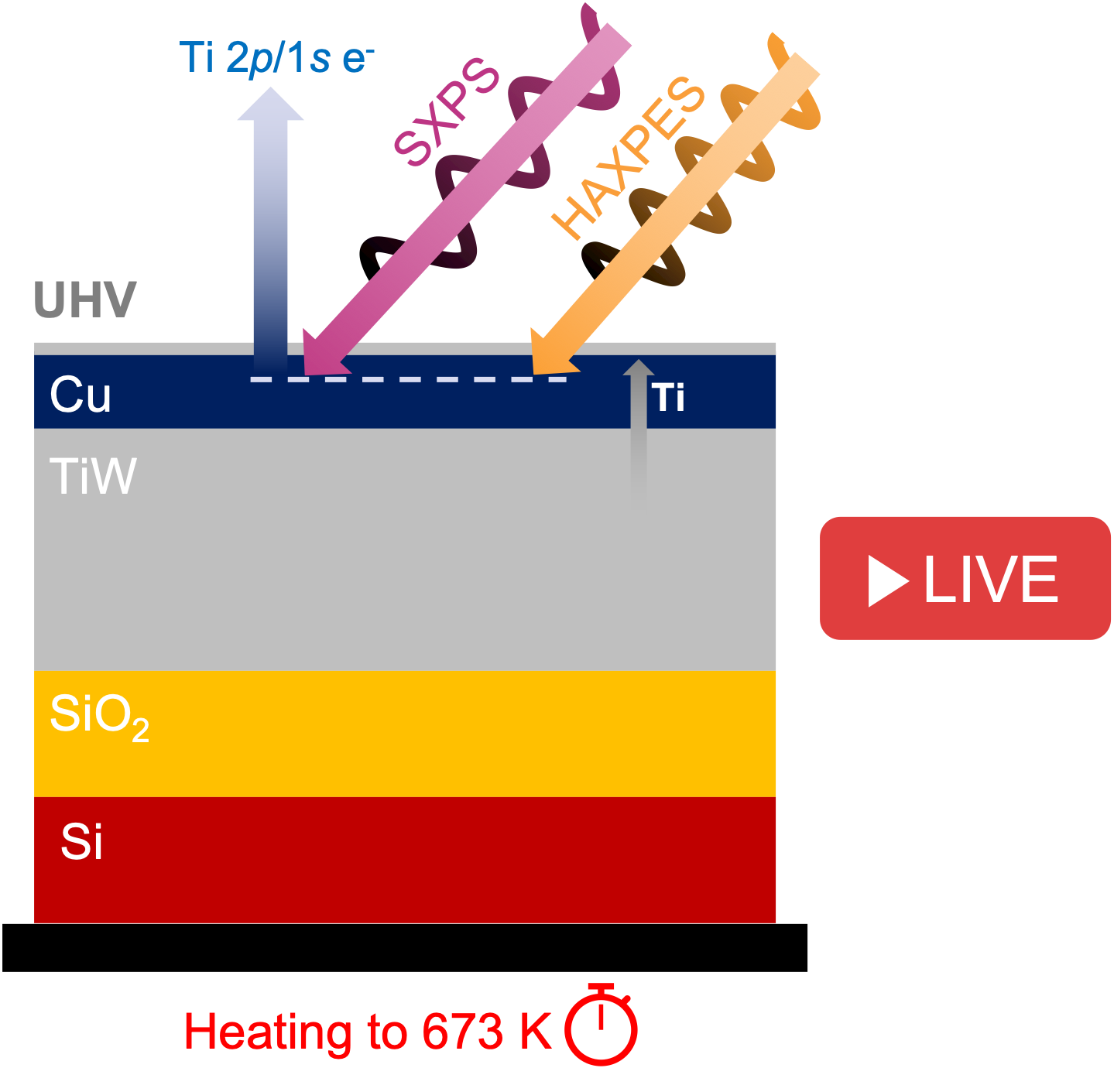}
  \medskip
  \caption*{The binary alloy of TiW is an attractive diffusion barrier for Si- and SiC-based power semiconductor devices that implement a copper metallisation scheme. However, at high temperatures, the barrier is found to degrade via the out-diffusion of Ti. This work explores the degradation mechanism using an in-situ X-ray photoelectron spectroscopy approach to monitor the diffusion in real-time.}
\end{figure}

\end{document}



\title{Capturing the dynamics of Ti diffusion across Ti$_x$W$_{1-x}$/Cu heterostructures using X-ray photoelectron spectroscopy}

\author{C.~Kalha}
\email{curran.kalha.19@ucl.ac.uk}
\affiliation{Department of Chemistry, University College London, 20 Gordon Street, London, WC1H~0AJ, United Kingdom.}

\author{P.~K.~Thakur}%
\author{T.-L.~Lee}%
\affiliation{Diamond Light Source Ltd., Diamond House, Harwell Science and Innovation Campus, Didcot, OX11 0DE, United Kingdom.}

\author{M.~Reisinger}
\author{J.~Zechner}
\author{M.~Nelhiebel}
\affiliation{Kompetenzzentrum Automobil- und Industrie-Elektronik GmbH, Europastraße 8, 9524 Villach, Austria.}

\author{A.~Regoutz}
 \email{a.regoutz@ucl.ac.uk}
\affiliation{Department of Chemistry, University College London, 20 Gordon Street, London, WC1H~0AJ, United Kingdom.}

\date{\today}

\maketitle

\newpage

 \tableofcontents

\cleardoublepage

\section{Peak fit analysis of as-deposited TiW spectra}

\begin{figure}[ht!]
\centering
    \includegraphics[keepaspectratio, width=0.8\linewidth]{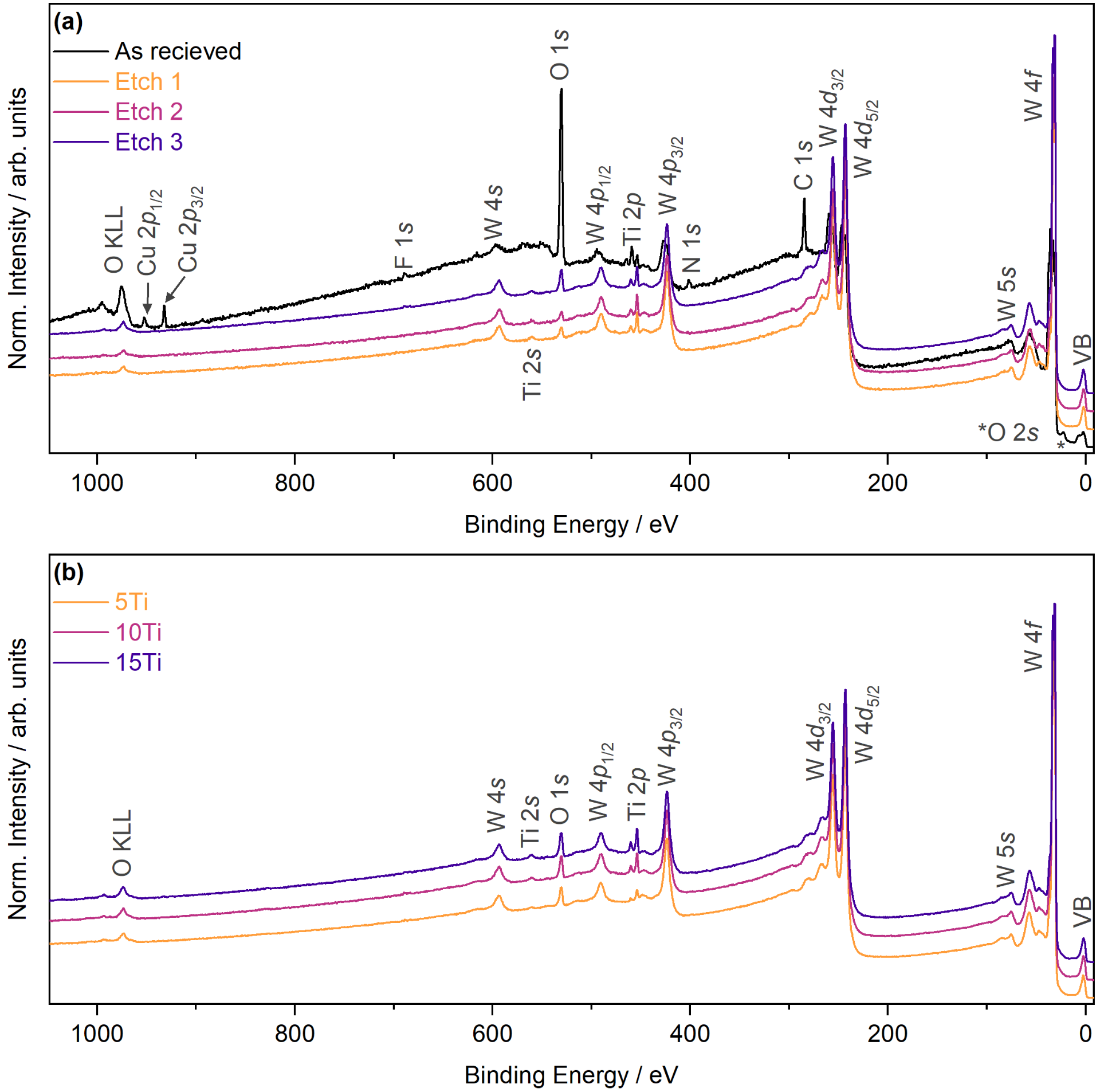}
    \caption{SXP survey spectra collected before and after in-situ preparation of samples, including (a) survey spectra collected for sample 10Ti after each etch step, and (b) survey spectra collected for all three samples at the end of the in-situ preparation method. Spectra are normalised (0-1) to the height of the most intense peak and are vertically offset. VB = valence band.}
    \label{fig:Sputter_Surv}
\end{figure}

To determine the Ti:W ratio of the as-deposited samples, the Ti~2\textit{p} and W~4\textit{f} core level spectra were collected with laboratory-based SXPS, after both ex-situ and in-situ preparation of the Si/SiO\textsubscript{2}/TiW/Cu samples. Samples were first cleaved to 5$\times$5~mm\textsuperscript{2} pieces using a diamond-tipped pen, after which they were submerged in a dilute solution of HNO\textsubscript{3} (5:1 65~\% conc. HNO\textsubscript{3}: Milli-Q water) for 10~min. This was carried out to selectively remove the copper metallisation layer without affecting the TiW layer. The samples were then sputter cleaned in-situ to remove contamination during the ex-situ preparation stages and any oxide formation. The survey spectra collected after the in-situ preparation are displayed in Fig.~\ref{fig:Sputter_Surv}.\par 

\begin{figure}[ht!]
\centering
    \includegraphics[keepaspectratio, width=0.6\linewidth]{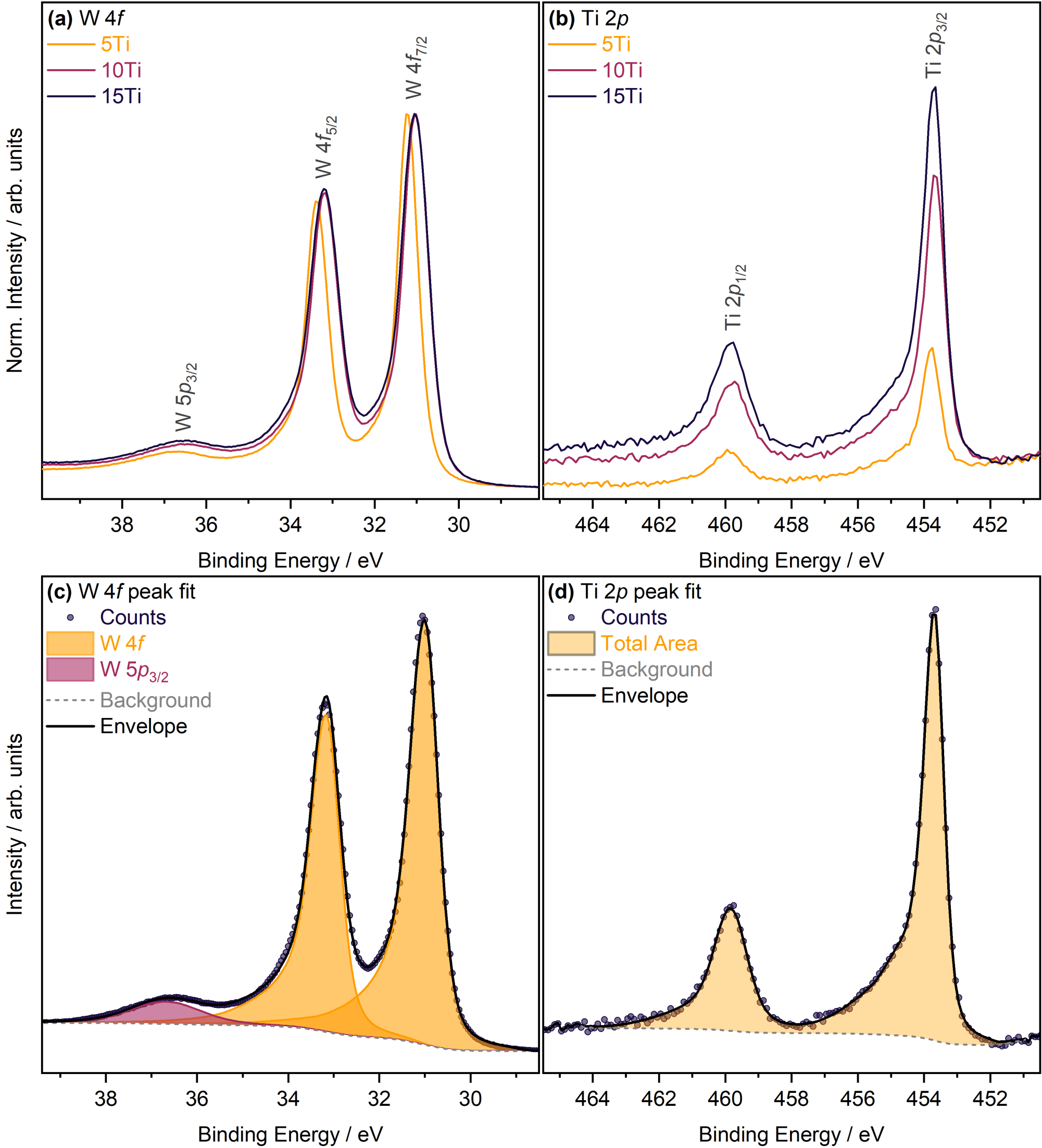}
    \caption{SXP core level spectra collected for all samples after the in-situ removal of the copper capping layer and oxide layer, and representative peak fits of the (a) W~4\textit{f} and (b) Ti~2\textit{p} core level spectra. Spectra are normalised to the W~4\textit{f}\textsubscript{7/2} peak height of the respective sample. Peak fits of the W~4\textit{f} and Ti~2\textit{p} core levels for spectra collected on the 10Ti sample are displayed in (c) and (d), respectively.}
    \label{fig:W4f_Ti2p}
\end{figure}

Once the sputter cleaning was performed, a depth profile using a focused Ar\textsuperscript{+} source was then conducted for each sample to determine the Ti:W concentration profile across the film. The depth profile consisted of six etching cycles, each lasting for 30~min while the Ar\textsuperscript{+} ion gun operated at a 500~eV accelerating voltage and 10~mA emission current. After six etch steps, the SiO\textsubscript{2} layer was detectable. The Ti~2\textit{p} and W~4\textit{f} core level spectra were collected at each etch step. Representative Ti~2\textit{p} and W~4\textit{f} spectra, along with representative peak fits, are displayed in Fig~\ref{fig:W4f_Ti2p}. Spectra were aligned to the intrinsic Fermi energy ($E_F$) of the respective sample. A systematic shift toward higher binding energy (BE) is observed in the W~4\textit{f} spectra with decreasing Ti, a trend that is also observed in the Ti~2\textit{p} spectra. \par

\begin{figure}[ht!]
\centering
    \includegraphics[keepaspectratio, width=\linewidth]{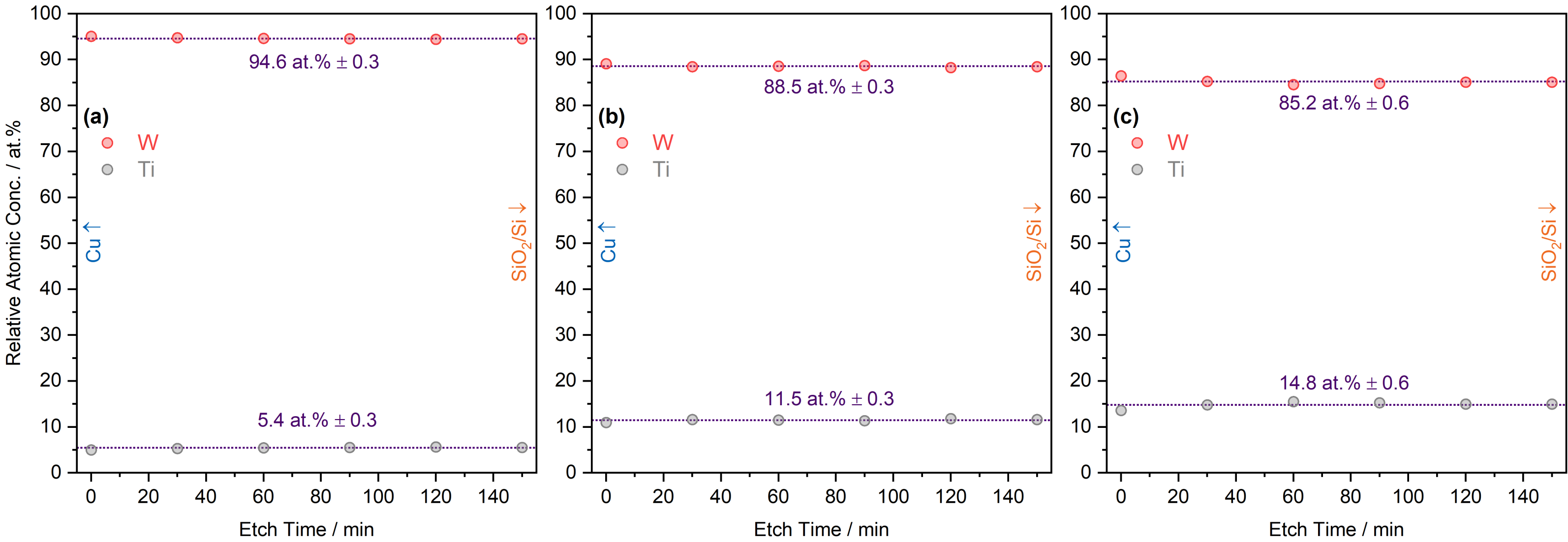}
    \caption{Ti:W relative quantification as a function of etching time (also referred to as sputter duration) across the three TiW films. The depth of profile of samples 5Ti, 10Ti and 15Ti is displayed in (a), (b), and (c), respectively. 0~min etch time refers to the first measured point in the depth profile. This was collected after the sample surface was in-situ sputter cleaned to remove the remnants from the ex-situ cleaning process but before the first etching cycle of the depth profile. This measurement point is referred to as Etch 0.}
    \label{fig:DP_Quant}
\end{figure}

\begin{figure}[ht!]
\centering
    \includegraphics[keepaspectratio, width=0.66\linewidth]{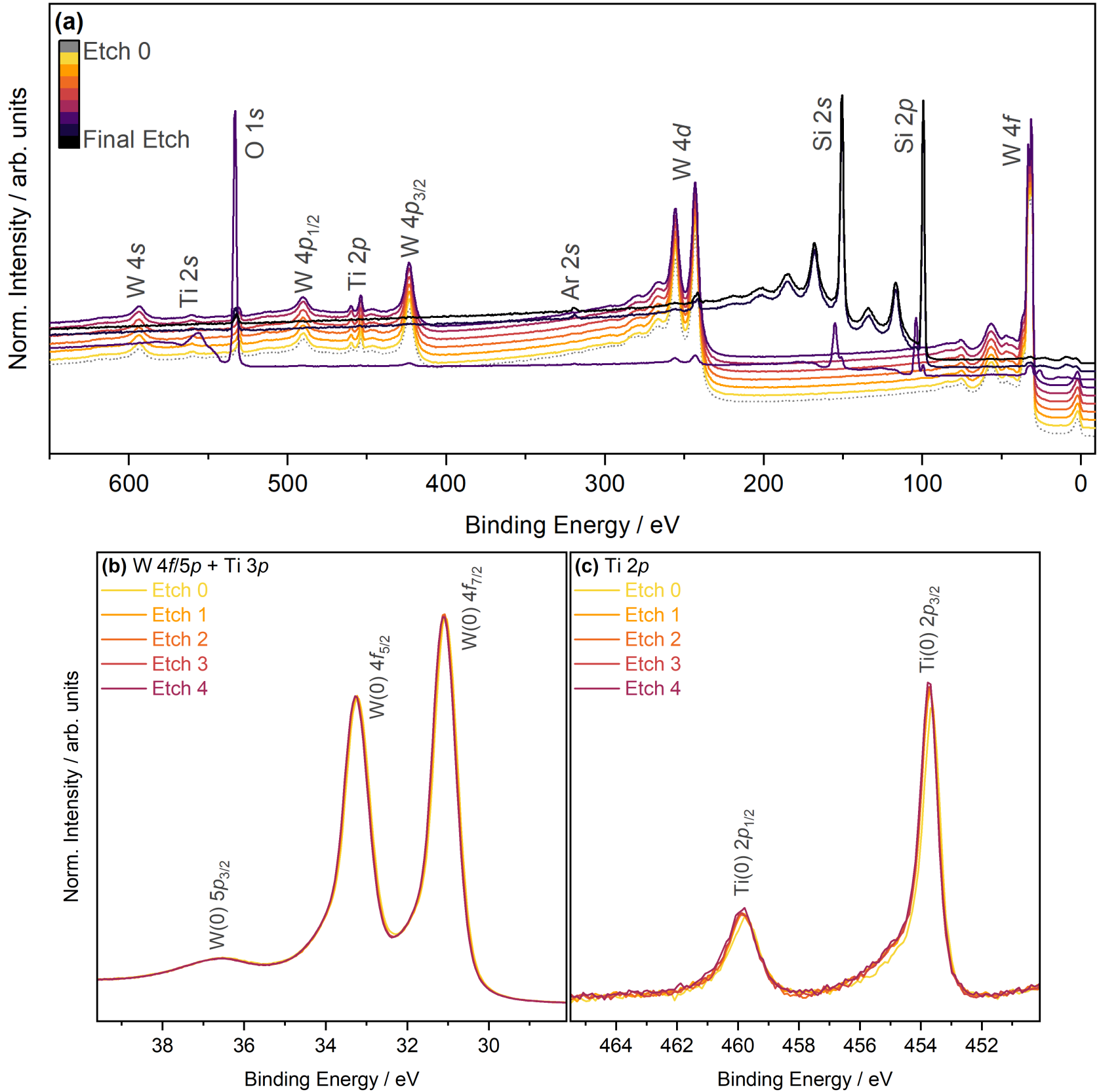}
    \caption{Spectra collected during the first five etch steps of the depth profile for sample 10Ti, including (a) survey, (b) W~4\textit{f}, and (c) Ti~2\textit{p} spectra. The survey spectra are normalised to the height of the maximum intensity peak, whereas the W~4\textit{f} and Ti~2\textit{p} spectra are normalised to the sum of the total W~4\textit{f}/5\textit{p}\textsubscript{3/2} and Ti~2\textit{p} areas. The dotted grey line in the survey spectra refers to the Etch 0 spectrum, and the survey spectra have been offset vertically. Etch 0 refers to the first measurement at sputtering time 0~min (i.e. after the in-situ cleaning but before the first depth profile etching cycle). As no Fermi edge or C~1\textit{s} was measured during the depth profiles, the BE scale is not calibrated and is plotted as recorded.}
    \label{fig:01_DPS}
\end{figure}

To determine the Ti:W ratio the Ti~2\textit{p} core level spectra collected across the entire depth profile were first fitted with the Smart-type background implemented in the Avantage software package, which is a Shirley-type background with the additional constraint that the background should not be greater than the data points. The smart background was chosen because at lower Ti concentrations, the background on the lower binding energy (BE) side of the Ti~2\textit{p} begins to rise due to the increase in intensity of the close neighbouring W~4\textit{p}\textsubscript{3/2} plasmon and this hampers the effective use of the Shirley-type background, as it would cut the data points. Due to the complexity of the Ti~2\textit{p} core level, the total area was fitted rather than to isolate the contributions from the two spin states. The average Ti~2\textit{p} relative atomic sensitivity factor (RASF) was applied to the resultant fitted area to quantify the region. For W~4\textit{f} the Shirley-type background was implemented and three peaks were added for the W~4\textit{f}\textsubscript{7/2}, W~4\textit{f}\textsubscript{5/2} and W~5\textit{p}\textsubscript{3/2} core lines. It is assumed that after sputtering only the metallic tungsten environment is present. The W~4\textit{f} peaks were given asymmetry to account for the core-hole coupling with conduction band states and constrained to have the same full width at half maximum (FWHM) and line shape as each other.~\cite{HUFNER1975417} The Avantage software package uses a least square fitting procedure to determine a suitable Lorentzian/Gaussian (L/G) mix, tail mix, full width at half maximum (FWHM), and tail exponent of the peaks. Additionally, the area ratio of the 4\textit{f} doublet peaks was set so that the lower spin state peak had an area that was 0.75 that of the higher spin state peak (i.e. 3:4 area ratio). The same line shape (FWHM, L/G mix, tail mix, tail exponent and area ratio) was applied to all W~4\textit{f} spectra across the depth profile. Additionally, the W~5\textit{p}\textsubscript{3/2} peak was fitted with a psuedo-Voigt profile peak with a fixed L/G mix of 30\% Lorentzian and a variable FWHM constraint. The BE range of the backgrounds, the line shapes, and FWHM constraints of the peaks was then applied to all spectra to be consistent across the sample set and the depth profiles. However, if the line shape was not constrained, the same value within error ($\pm$0.3~at.\%) was achieved. To determine the relative Ti:W ratio in at.\%, the RASF corrected Ti~2\textit{p} spectral area was compared to the RASF corrected W~4\textit{f}\textsubscript{7/2} spectral area. Fig.~\ref{fig:DP_Quant} displays the quantification results from the depth profiles along with a standard deviation across the film thickness. The three samples have an average Ti~at.\% relative to W of 5.4$\pm$0.3 (5Ti), 11.5$\pm$0.3 (10Ti) and 14.8$\pm$0.6~at.\% (15Ti). Furthermore, Fig.~\ref{fig:01_DPS} displays the spectra collected across the depth profile of sample 10Ti, and it can be seen that the W~4\textit{f} line shape remains fairly constant across the first five etch steps, and subtle changes are observed in the W~4\textit{f}/Ti~2\textit{p} area ratio, reflecting what is observed with the values from the quantification. The survey spectra displayed in Fig.~\ref{fig:01_DPS}(a) also nicely show how the depth profile penetrates across the TiW and into the substrate, as in the last three etches, Si-O peaks first emerge, followed by Si peaks.

\cleardoublepage




\section{Room temperature energy resolution}

\begin{figure}[ht!]
\centering
    \includegraphics[keepaspectratio, width=0.4\linewidth]{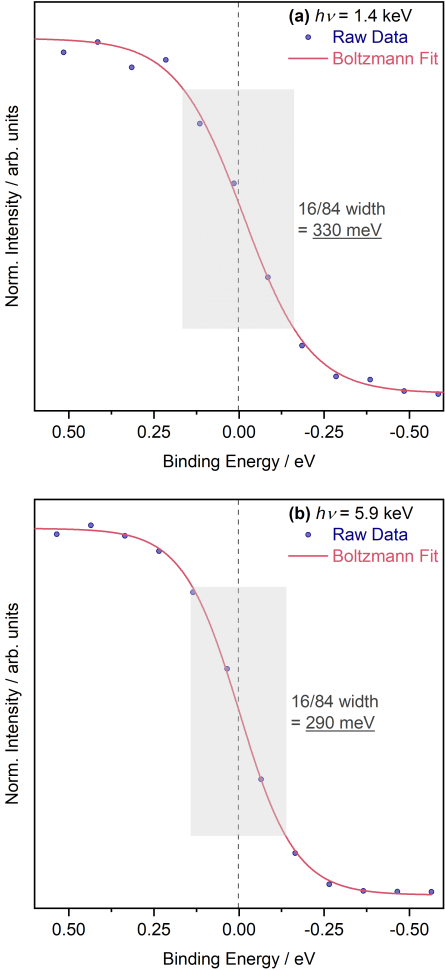}
    \caption{Fermi edge (E\textsubscript{F}) spectra collected with (a) SXPS and (b) HAXPES on a polycrystalline gold foil at room temperature. The energy resolution is determined by extracting the 16/84\% width (i.e.\ one standard deviation on either side of the Fermi energy.}
    \label{fig:Au_Ef}
\end{figure}

The room temperature total energy resolution of the SXPS and HAXPES experiments at the synchrotron was determined by determining the 16/84\% width of the Fermi edge of a polycrystalline gold foil. Fig.~\ref{fig:Au_Ef} displays the Fermi edges of the foil measured with SXPS and HAXPES at room temperature and fitted with a Boltzmann curve.

\cleardoublepage

\section{Sample Plate Holder}

\begin{figure}[ht!]
\centering
    \includegraphics[keepaspectratio, width=0.4\linewidth]{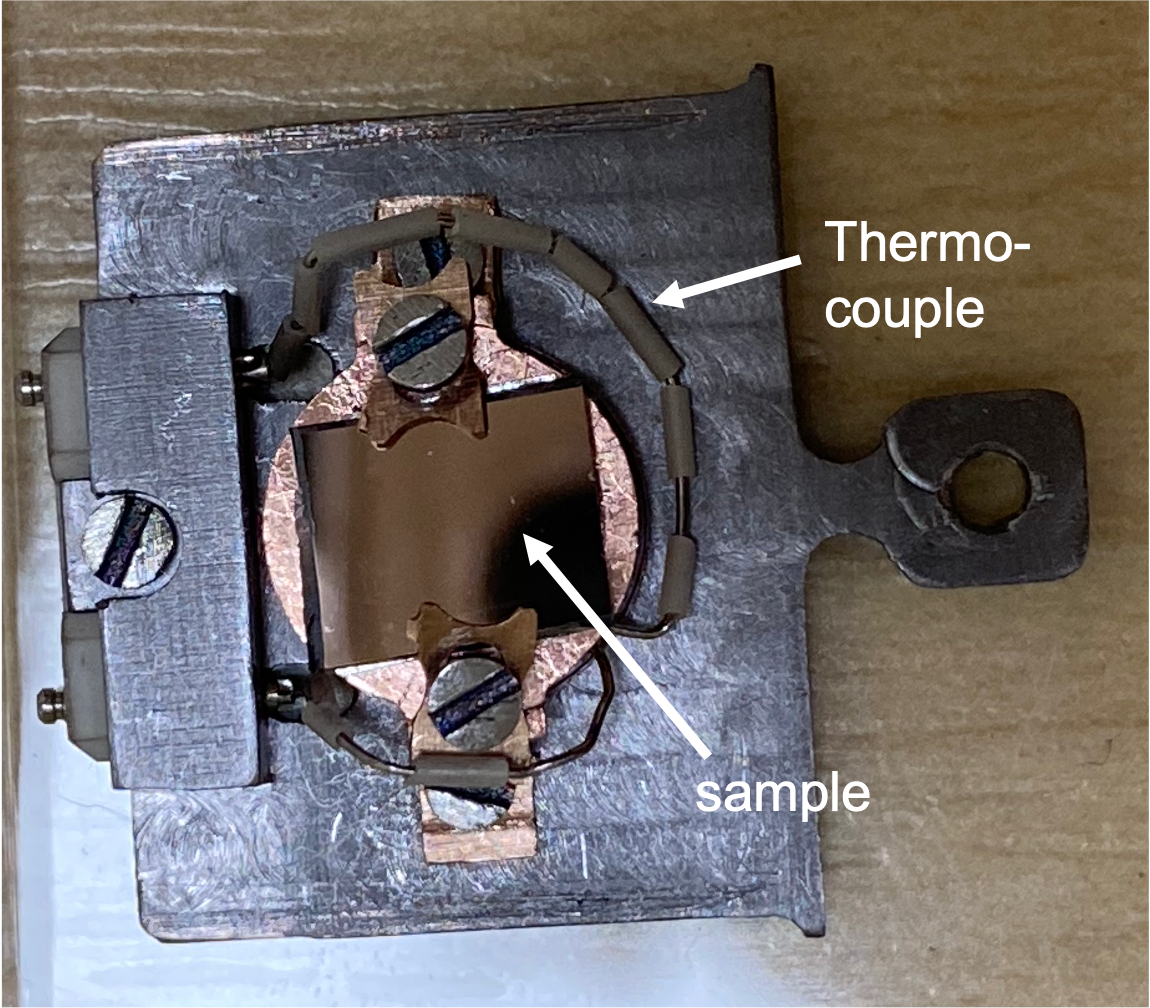}
    \caption{Annotated image of the sample plate holder used for the in-situ annealing experiment at beamline I09.}
    \label{fig:Sample_Plate}
\end{figure}

\cleardoublepage

\section{Temperature Profiles}

\begin{figure}[ht!]
\centering
    \includegraphics[keepaspectratio, width=0.42\linewidth]{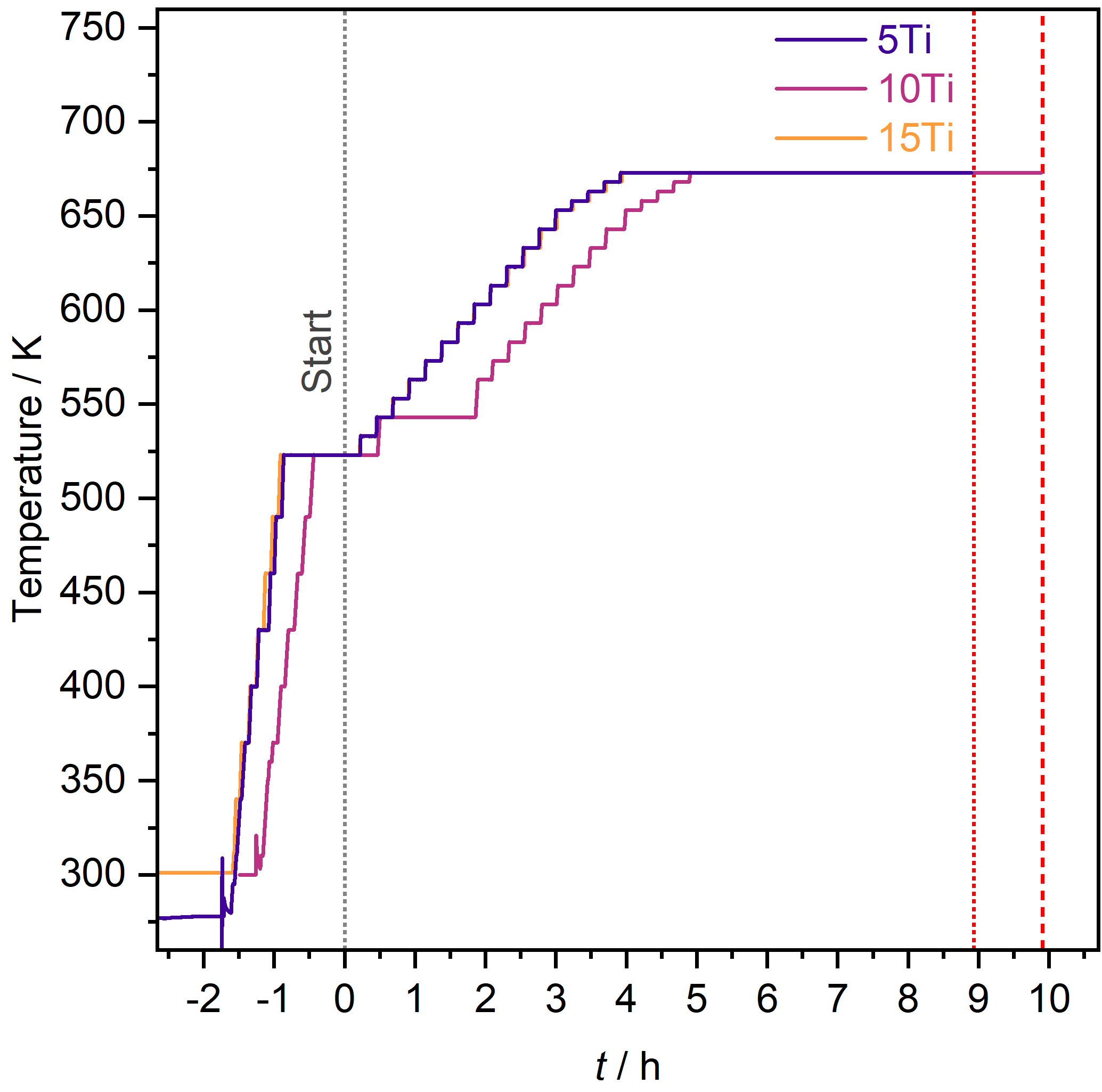}
    \caption{Temperature profiles for all three samples. The start of the measurement window is indicated by the vertically dotted grey line, whereas the red dotted and dashed lines indicate the end of the measurement cycle for samples 5Ti/15Ti and 10Ti, respectively. The temperature profile for samples 5Ti and 10Ti are near-identical and so overlap.}
    \label{fig:T_profile}
\end{figure}

\cleardoublepage

\section{Energy resolution as a function of temperature}

\begin{figure}[ht!]
\centering
    \includegraphics[keepaspectratio, width=\linewidth]{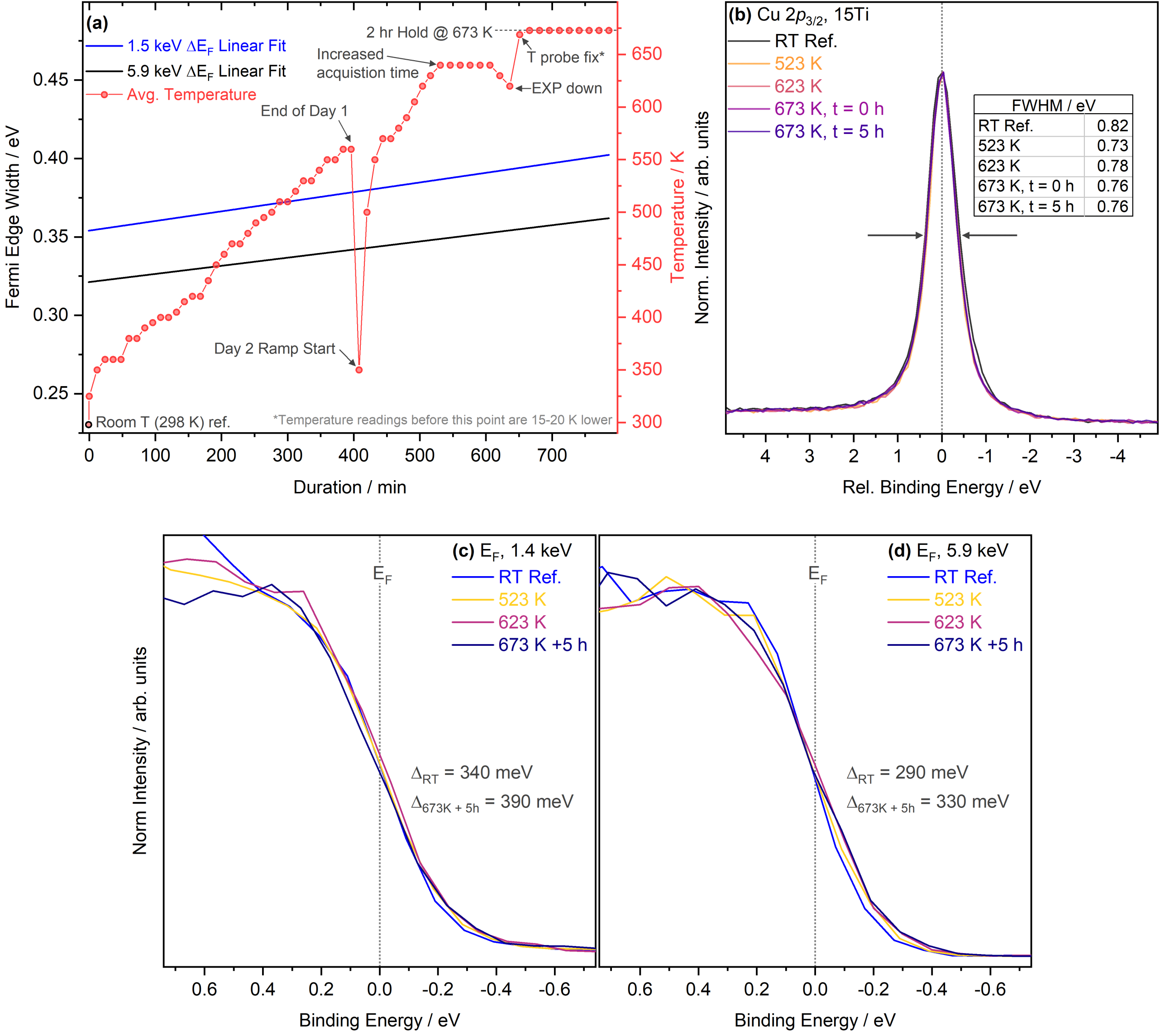}
    \caption{Energy resolution measurements as a function of annealing temperature and duration, including (a) the Fermi edge width collected with both soft (SX) and hard (HX) X-rays for sample 10Ti as a function of temperature during preliminary measurements, (b) selected Cu~2\textit{p}\textsubscript{3/2} core level spectra collected with SXPS on sample 15Ti as a function of annealing temperature, collected during this experiment, plotted on a relative BE scale and normalised to the maximum intensity to emphasis the change in peak FWHM as a function of annealing temperature and duration. (c) and (d) display the selected Fermi edge spectra collected as a function of annealing temperature measured with soft and hard X-rays, respectively. (c) and (d) are normalised to the maximum height (accounting for noise) of the Fermi edge and plotted on the same \textit{y}-axis scale. RT Ref. refers to the room temperature reference spectrum.}
    \label{fig:Reso_T}
\end{figure}

In order to assess the effect of temperature on the thermal broadening of the collected spectra, the intrinsic Fermi edge of the sample (i.e. copper) was captured with SXPS and HAXPES at each spectral cycle. By extracting the 16/84\% width of the Fermi edge (as shown in Fig.~\ref{fig:Au_Ef}), the change in total energy resolution could be monitored with respect to temperature. According to M\"{a}hl~\textit{et al.} the thermal broadening ($\gamma_f$) of a Fermi edge at temperature $T$ measured with XPS can be described by:

\begin{equation}
    {\gamma}_f = 4{\ln}(\sqrt{2}+1)k_b{T}\;{\approx}\;{\frac{7}{2}}k_b{T} ,
\end{equation}
where $k_b$ is the Boltzmann constant and approximating ${k_b}T$ to $\frac{T}{11600}\frac{\textrm{eV}}{\textrm{K}}$ gives a value of 90~meV and 200~meV for the thermal broadening at 300~K and 673~K, respectively.~\cite{MAHL1997197} Therefore, a change of 110~meV in the total energy resolution of this experiment is expected. Fig.~\ref{fig:Reso_T}(a) displays the change in Fermi edge width with respect to annealing temperature and duration during preliminary test measurements.\par

It can be seen in Fig.~\ref{fig:Reso_T}(a) that across the measured temperature range, on average the change in 16/84\% Fermi edge width is less than 60~meV. Considering everything remains constant during the measurement (i.e. pass energy, dwell time, analyser, geometry, sample) except for temperature, this change is representative of the thermal broadening. This value is slightly lower than the theoretical value, but this can be attributed to the assumptions made in the theoretical model and the error associated with the 16/84\% method. Additionally, Fig.~\ref{fig:Reso_T}(c) and (d) display the Fermi edge spectra at key temperatures measured in this experiment for sample 15Ti. The changes observed are minimal, with the hard X-ray-collected Fermi edges appearing more sensitive to temperature than the soft X-ray-collected edges.\par

Overall, the change in resolution is insignificant for the core level spectra as it falls below the energy resolution of the spectrometer. Therefore, when analysing the changes to the core level spectra for all samples, thermal broadening effects are negligible. Moreover, Fig.~\ref{fig:Reso_T}(b) displays the Cu~2\textit{p}\textsubscript{3/2} core level spectrum collected at selected temperatures. The room temperature spectrum is slightly broader than the higher temperature spectra, but the high-temperature spectra FWHM remain reasonably constant, falling in line with the changes observed when tracking the Fermi edge width. The reason for the broader room temperature spectrum and slight asymmetry on the lower binding energy side can be attributed to surface contamination (i.e. remnant oxide contributions) but when heated, the surface is cleaned, leading to a narrowing of the FWHM.

\cleardoublepage

\section{Room temperature reference spectra}

\begin{figure}[ht!]
\centering
    \includegraphics[keepaspectratio, width=0.7\linewidth]{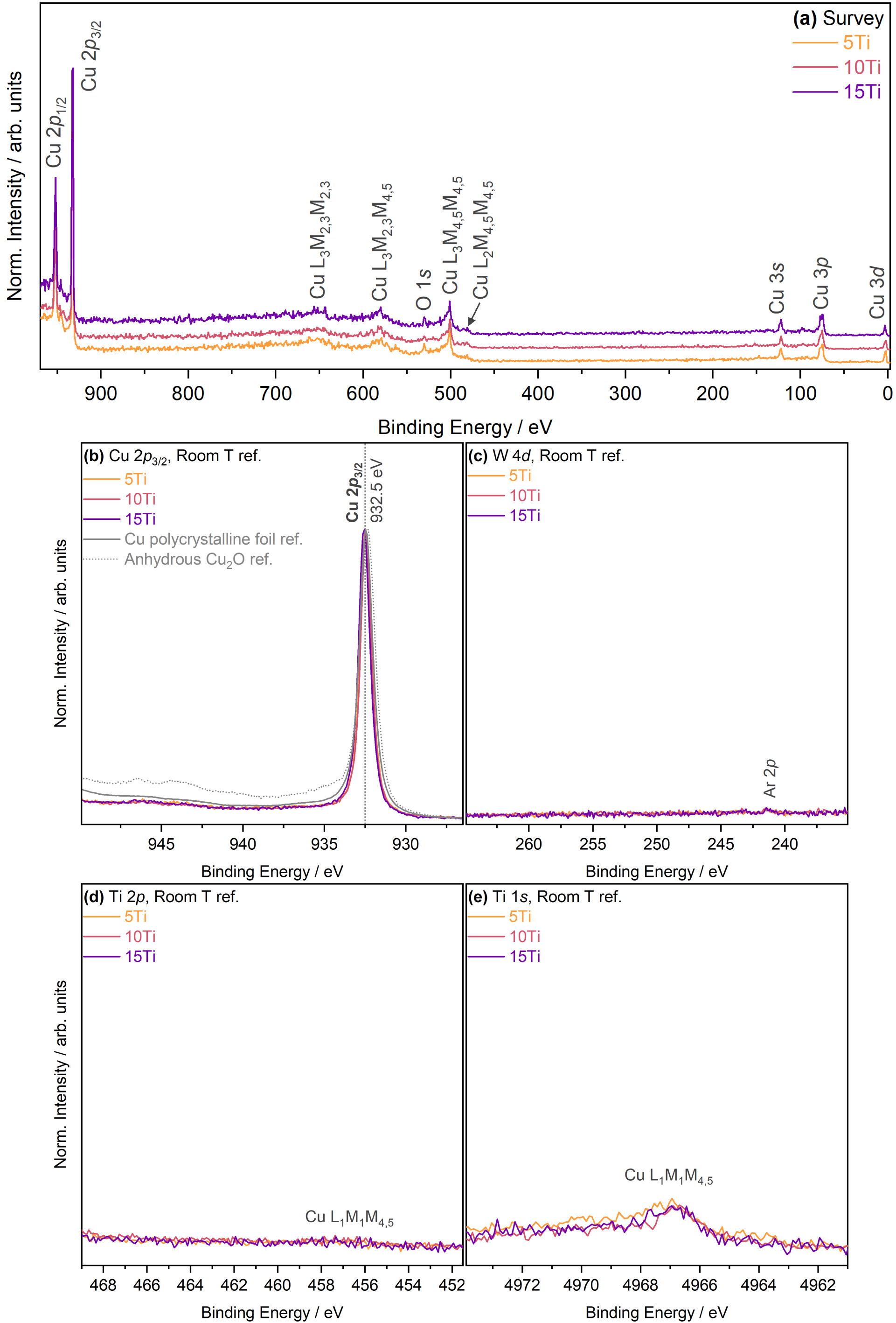}
    \caption{SXPS and HAXPES room-temperature reference spectra collected for as-deposited samples 5Ti, 10Ti and 15Ti after the surface was in-situ cleaned via argon sputtering, including (a) survey, (b) Cu~2\textit{p}\textsubscript{3/2}, (c) W~4\textit{d}, (d) Ti~2\textit{p} and (e) Ti~1\textit{s}, with the Ti~1\textit{s} collected with HAXPES and the others with SXPS. Spectra are normalised to the maximum height of the Cu~2\textit{p}\textsubscript{3/2} signal. Spectra collected on reference copper compounds (Cu, Cu\textsubscript{2}O) are also included, which were measured using the laboratory-based SXPS instrument.}
    \label{fig:Refs_RoomT}
\end{figure}

To have confidence in the interpretation of the Cu~2\textit{p}\textsubscript{2/3} spectra, reference measurements were conducted using laboratory-based SXPS instrument ($h\nu$ = 1.4867~keV) on a polycrystalline Cu foil (Alfa Aesar, 99.9985\% metals basis, 0.25~mm thick) and an anhydrous Cu\textsubscript{2}O powder (Cu\textsubscript{2}O, Sigma Aldrich, $>=$99.99\% metals basis). The foil reference was sputter cleaned in-situ using a focused argon ion beam and sputtering for 10~min, with the ion gun operating at 2~keV voltage. The Cu\textsubscript{2}O powder was received in a sealed ampule under an argon atmosphere, and to minimise further oxidation (i.e. the formation of CuO) the sample was prepared in a glovebag under argon. The recorded Cu~2\textit{p}\textsubscript{3/2} spectra of these reference materials are overlaid on the room temperature reference spectra of samples 5, 10 and 15Ti, displayed in Fig.~\ref{fig:Refs_RoomT}(b). The binding energy scale was calibrated to the intrinsic Fermi energy for the TiW/Cu samples and the Cu foil reference, whereas for Cu\textsubscript{2}O the scale was calibrated to adventitious carbon (284.8~eV).\par

It can be observed, that there is good agreement between the Cu foil reference and the spectra recorded for the TiW/Cu samples. A very weak satellite is observed between 942-948~eV for the TiW/Cu samples, however, this is also present in the Cu foil reference, therefore indicating that the native oxide contribution has been minimised as much as possible. The slight differences in Cu~2\textit{p}\textsubscript{3/2} FWHM between the foil reference and TiW/Cu samples can be explained by the differences in total energy resolution between the synchrotron ($h\nu$ = 1.4~keV) and laboratory-based measurements, which were determined to be 330~meV and 600~meV, respectively. The laboratory-based SXPS instrument used for the collection of reference spectra was not the same used for the depth profiles described in the manuscript, hence the different energy resolution. \par

Cu Auger peaks are identified to overlap with the measured Ti~2\textit{p} and Ti~1\textit{s} core levels when measured with h$\nu$ = 1.4 and 5.9~keV, respectively. The Auger peak appears at a BE position of $\approx$4967.0~eV in the Ti ~1\textit{s} region and $\approx$457.0~eV in the Ti~2\textit{p} region, equating to a kinetic energy (KE) of $\approx$959.0~eV for both the Auger peaks. The reason why they both have the same kinetic energy is due to the strategic decision to tune the photon energies so that the Ti~1\textit{s} and Ti~2\textit{p} probing depths match. Possible Auger transition energies have been calculated and tabulated by Coghlan~\textit{et al.},~\cite{COGHLAN1973317} and the position of the Auger in the Ti~1\textit{s} spectra correlates with the Auger Cu~L\textsubscript{1}M\textsubscript{1}M\textsubscript{4,5} transition calculated at 962~eV (KE). It is clear that these peaks are not due to titanium as they do not possess the attributes of a core level peak nor the expected BE position of titanium metal/oxide in either the 2\textit{p} or 1\textit{s} spectrum. Aside from the Cu Auger peaks, the Ar~2\textit{p} core level peak is visible in the W~4\textit{d} region at approximately 241.0~eV corresponding to implanted argon from the sputtering process. However, this peak is again incredibly small and does not affect the analysis of the W~4\textit{d} spectrum that may develop during annealing.

\cleardoublepage

\section{In-situ annealing Ti~2\textit{p}~core level spectra}

\begin{figure}[ht!]
\centering
    \includegraphics[keepaspectratio, width=\linewidth]{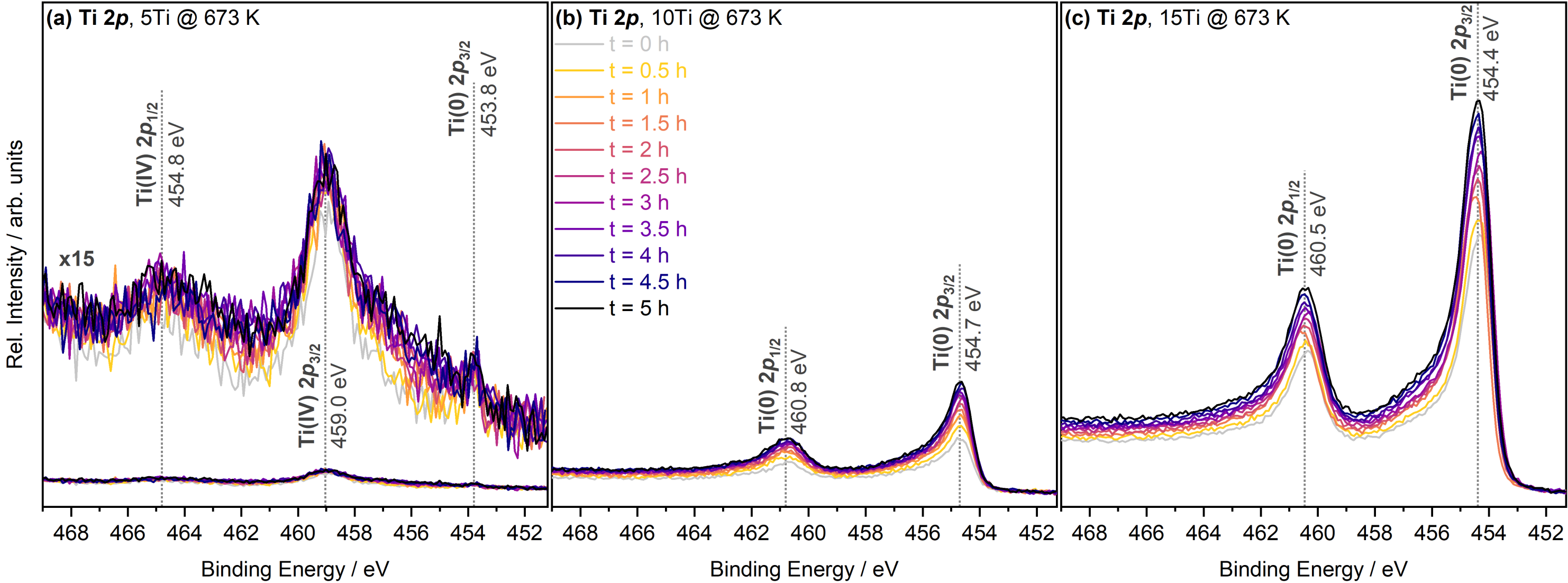}
    \caption{Ti~2\textit{p} core level spectra collected during the 673~K holding period (Stage \textbf{3}) for sample (a) 5Ti, (b) 10Ti, and (c) 15Ti. Spectra for each core level are plotted over the same $y$-axis scale to show the differences in intensity across the three samples. The spectra have not been normalised but a constant linear background has been removed. Additionally, spectra recorded every other spectral cycle are displayed to aid with the interpretation of the data. The 5Ti spectra have been magnified by $\times$15 to aid with viewing. The legend displayed in (b) also applies to (a) and (c). Ti(0) and Ti(IV) refers to metallic Ti and titanium oxide in the 4+ oxidation state, respectively.}
    \label{fig:Ti2p_core_levels}
\end{figure}

\cleardoublepage

\section{Heat map of Ti~1\textit{s} spectra collected over the measurement window}

\begin{figure}[ht!]
\centering
    \includegraphics[keepaspectratio, width=\linewidth]{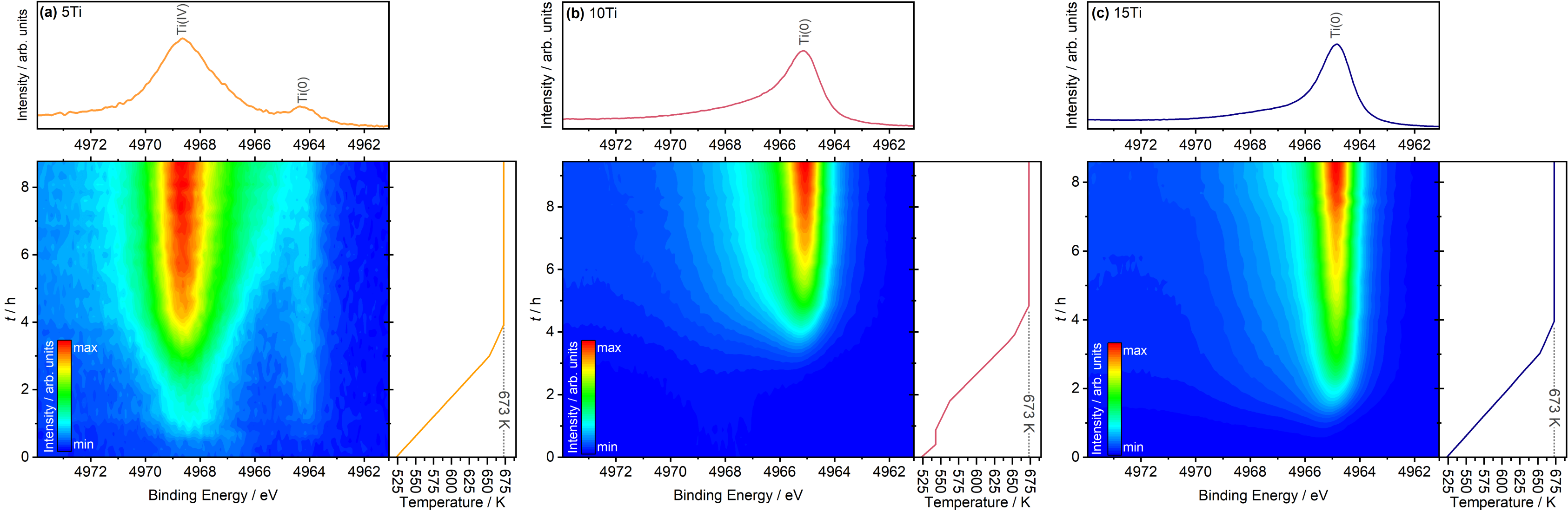}
    \caption{HAXPES maps of the Ti~1\textit{s} core level collected across the entire measurement window, for sample (a) 5Ti, (b) 10Ti and (c) 15Ti. The spectra are aligned to the intrinsic Fermi energy of the respective sample, and their intensity is not normalised but plotted as-collected (after the subtraction of a constant linear background). The top panel displays the median spectrum collected across the measurement window and the right panel displays the point-by-point temperature profile as a function of time. Due to the large variation in spectral intensity between sample 5Ti and 15Ti, the spectra displayed here are on independent intensity scales and so the intensities should not be directly compared. Ti(0) and Ti(IV) refers to metallic Ti and titanium oxide in the 4+ oxidation state, respectively.}
    \label{fig:Ti1s_heat}
\end{figure}

\cleardoublepage

\section{5Ti Ti~1\textit{s} peak fit analysis}

\begin{figure}[ht!]
\centering
    \includegraphics[keepaspectratio, width=0.4\linewidth]{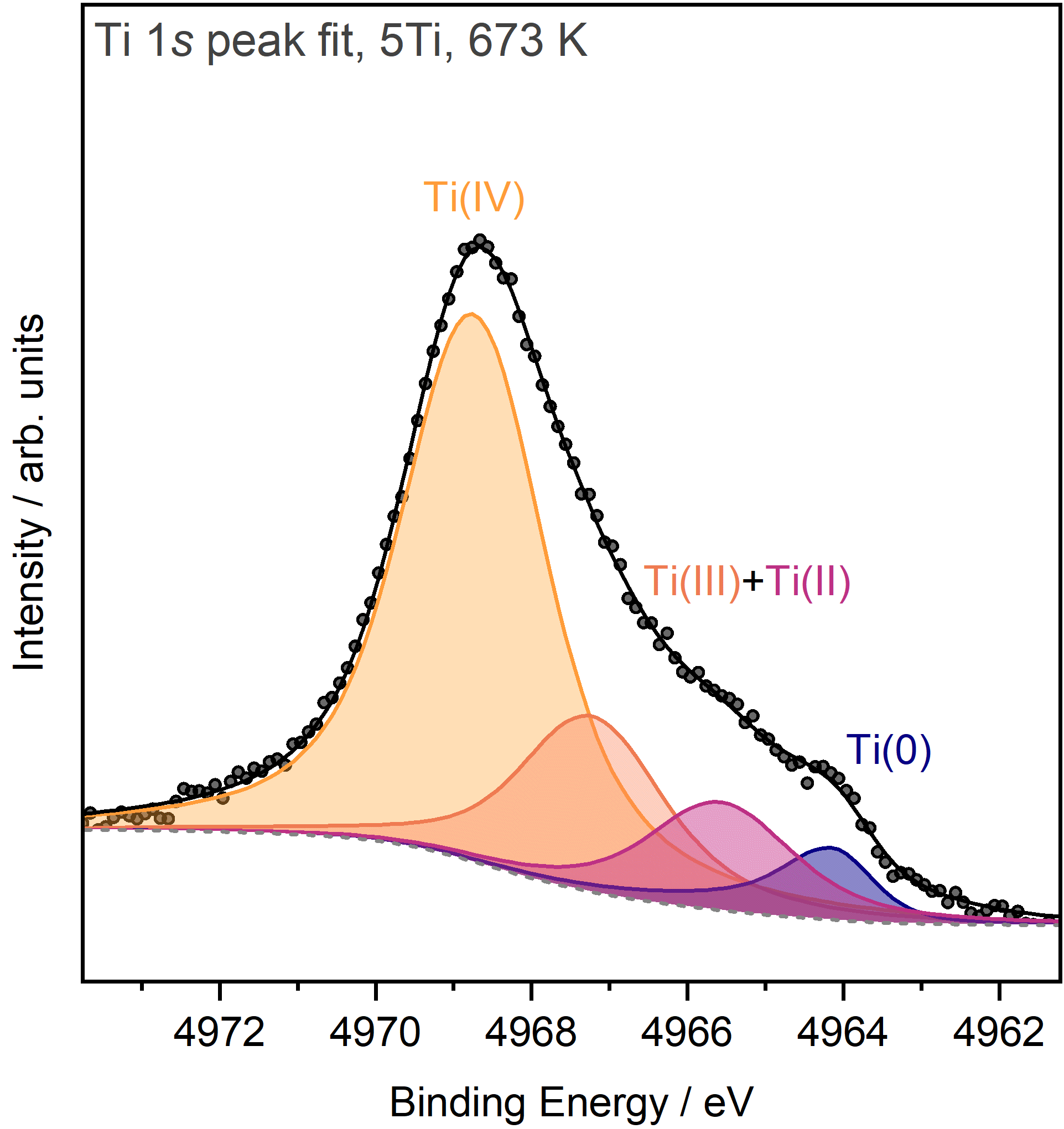}
    \caption{Peak fit analysis of the Ti~1\textit{s} core level for sample 5Ti. The oxide peaks are constrained to have the same FWHM (2.2~eV) and Lorentzian/Gaussian mix (50), whereas the metal peak line shape was derived from peak fitting the 673~K spectra of sample 30Ti with one asymmetric line shape. A Shirley-type background was used, and the Cu~L\textsubscript{1}M\textsubscript{1}M\textsubscript{4,5} contribution was not removed.}
    \label{fig:Ti1s_pf_5Ti}
\end{figure}

\cleardoublepage
\section{Residual oxygen within the as-deposited Cu film}

\begin{figure}[ht!]
\centering
    \includegraphics[keepaspectratio, width=\linewidth]{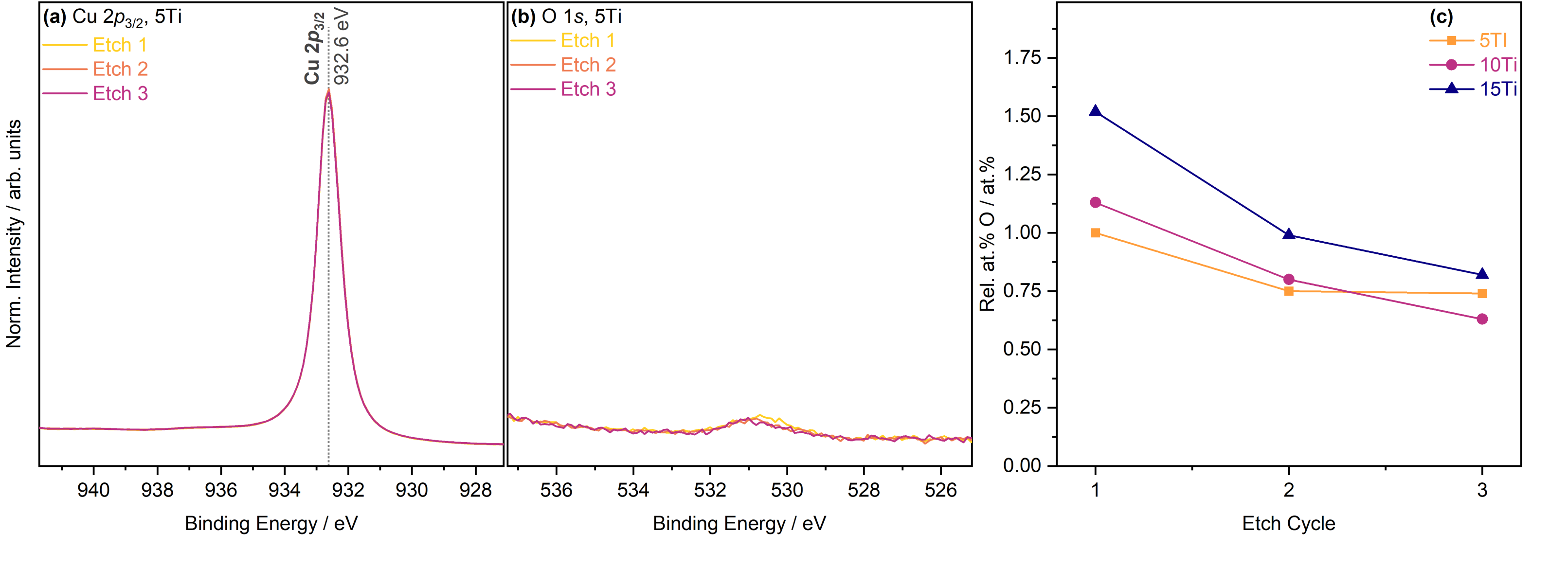}
    \caption{Depth profile results across the three as-deposited TiW/Cu samples to determine the level of O within the bulk Cu film. Samples were sputtered using a focused 500~eV Ar\textsuperscript{+} ion-beam gun energy for 6 min, rastering over a 2$\times$2~mm\textsuperscript{2} area and measuring at the centre of the sputter crater. Three cycles of sputtering were conducted equating to 18 min total sputtering time. (a) and (b) show the Cu~2\textit{p}\textsubscript{3/2} and O~1\textit{s} spectra collected after the first, second and third etch steps for sample 5Ti only, respectively. Etch 0 refers to the as-received measurement (i.e. before any sputtering) and is not included here as the samples were stored and handled in air so a thin native oxide and adventitious carbon layer were present. The quantification results of the O/(Cu+O) ratio at each of the three etch steps for all three samples are shown in (c). The spectra are aligned to the ISO standard BE value of metallic Cu~2\textit{p}\textsubscript{3/2} (932.62~eV)~\cite{Cu_ISO} and normalised to the Cu~2\textit{p}\textsubscript{3/2} total spectral area. After Etch 3, the TiW layer is reached and the Ti and W signals become dominant.}
    \label{fig:O_traces}
\end{figure}

\cleardoublepage



\section{Early Stages of Annealing for Sample 5Ti}

\begin{figure}[ht!]
\centering
    \includegraphics[keepaspectratio, width=0.67\linewidth]{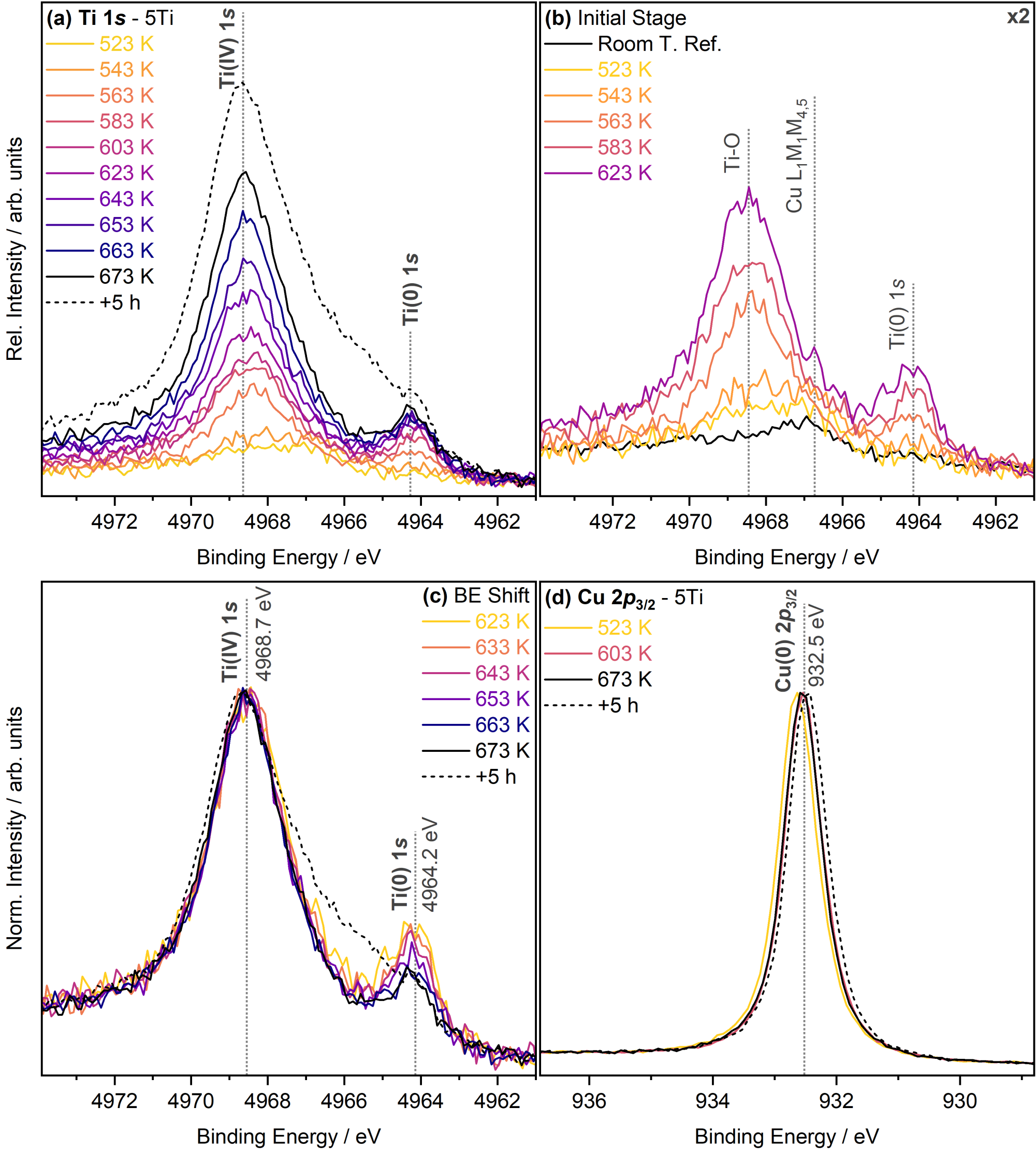}
    \caption{Initial stages of annealing (523-673~K) described by the Cu~2\textit{p}\textsubscript{3/2} and Ti~1\textit{s} core level spectra for sample 5Ti. (a) Ti~1\textit{s} core level spectra collected (with no intensity normalisation) at each temperature increment, with +5~h referring to the data collected at the end of the 5~h 673~K holding period. (b) A magnified view of the Ti~1\textit{s} core level spectra collected between 523-623~K as well as a room temperature reference measurement on the same sample (prior to annealing) to highlight the Cu Auger contribution. (c) Normalised (0-1) Ti~1\textit{s} core level spectra to emphasise the change in line shape. (d) Normalised (0-1) Cu~2\textit{p}\textsubscript{3/2} spectra taken at selected temperatures. All data have been aligned to the intrinsic Fermi energy. (a) and (b), and (c) and (d) are plotted on the same $y$-axis scale.}
    \label{fig:5Ti_early}
\end{figure}

\cleardoublepage
\section{Early Stages of Annealing for Sample 15Ti}

\begin{figure}[ht!]
\centering
    \includegraphics[keepaspectratio, width=0.67\linewidth]{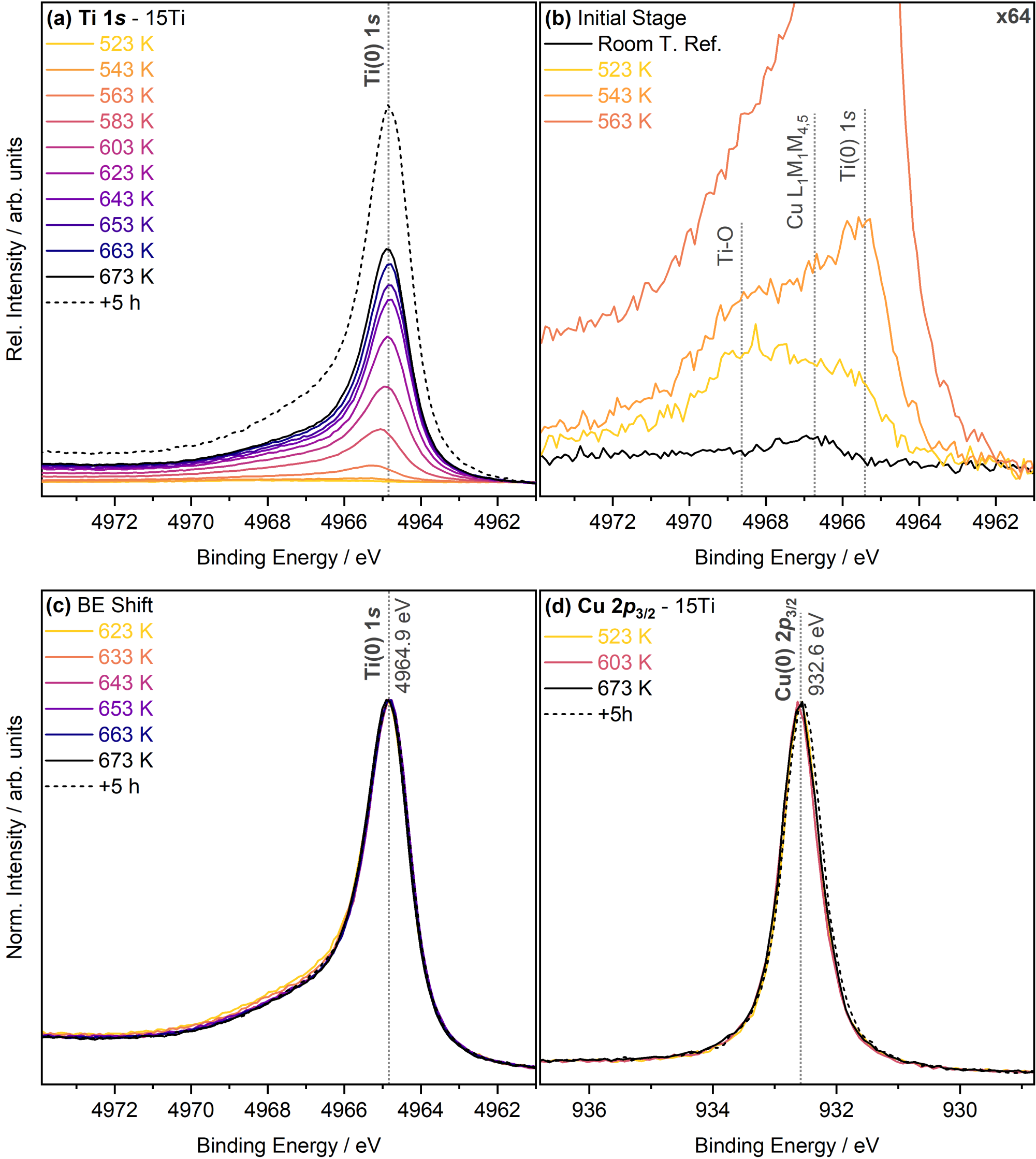}
    \caption{Initial stages of annealing (523-673~K) described by the Cu~2\textit{p}\textsubscript{3/2} and Ti~1\textit{s} core level spectra for sample 15Ti. (a) Ti~1\textit{s} core level spectra collected (with no intensity normalisation) at each temperature increment, with +5~h referring to the data collected at the end of the 5~h 673~K holding period. (b) A magnified view of the Ti~1\textit{s} core level spectra collected between 523-623~K as well as a room temperature reference measurement on the same sample (prior to annealing) to highlight the Cu Auger contribution. (c) Normalised (0-1) Ti~1\textit{s} core level spectra to emphasise the change in line shape. (d) Normalised (0-1) Cu~2\textit{p}\textsubscript{3/2} spectra taken at selected temperatures. All data have been aligned to the intrinsic Fermi energy. (a) and (b), and (c) and (d) are plotted on the same $y$-axis scale.}
    \label{fig:15Ti_early}
\end{figure}

\cleardoublepage

\section{Cu~2\textit{p}\textsubscript{3/2} line shape changes}

\begin{figure}[ht!]
\centering
    \includegraphics[keepaspectratio, width=0.4\linewidth]{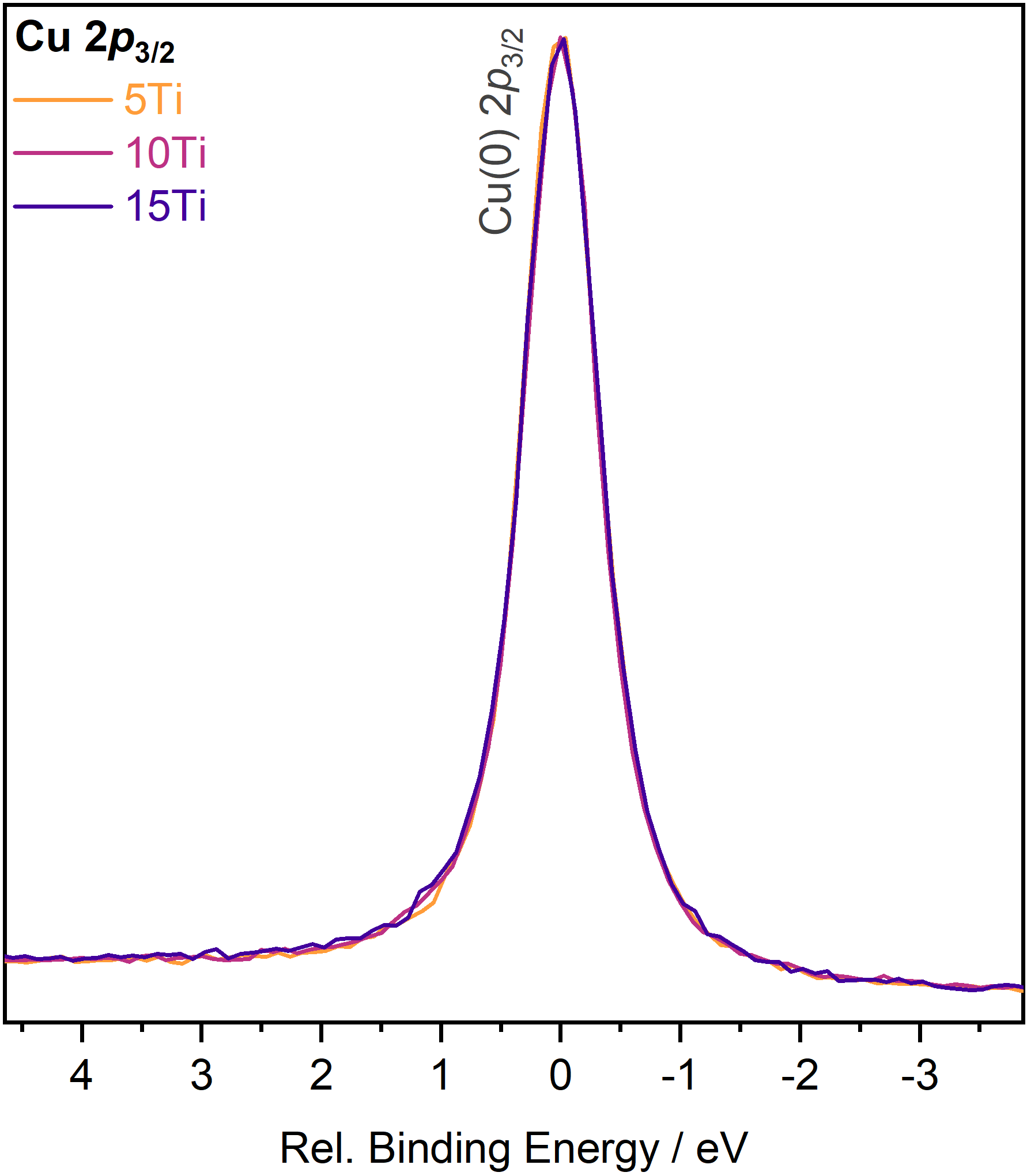}
    \caption{Comparison of the Cu~2\textit{p}\textsubscript{3/2} spectral line shape of the three samples. The spectra presented were captured at the end of the 673~K holding period (i.e. 673~K + 5~h). The spectra are normalised 0-1 and aligned to the main intensity to make it easier to observe changes in the line shape.}
    \label{fig:Cu2p}
\end{figure}

\cleardoublepage

\section{In-situ annealing Ti~2\textit{p} concentration profile}

\begin{figure}[ht!]
\centering
    \includegraphics[keepaspectratio, width=0.4\linewidth]{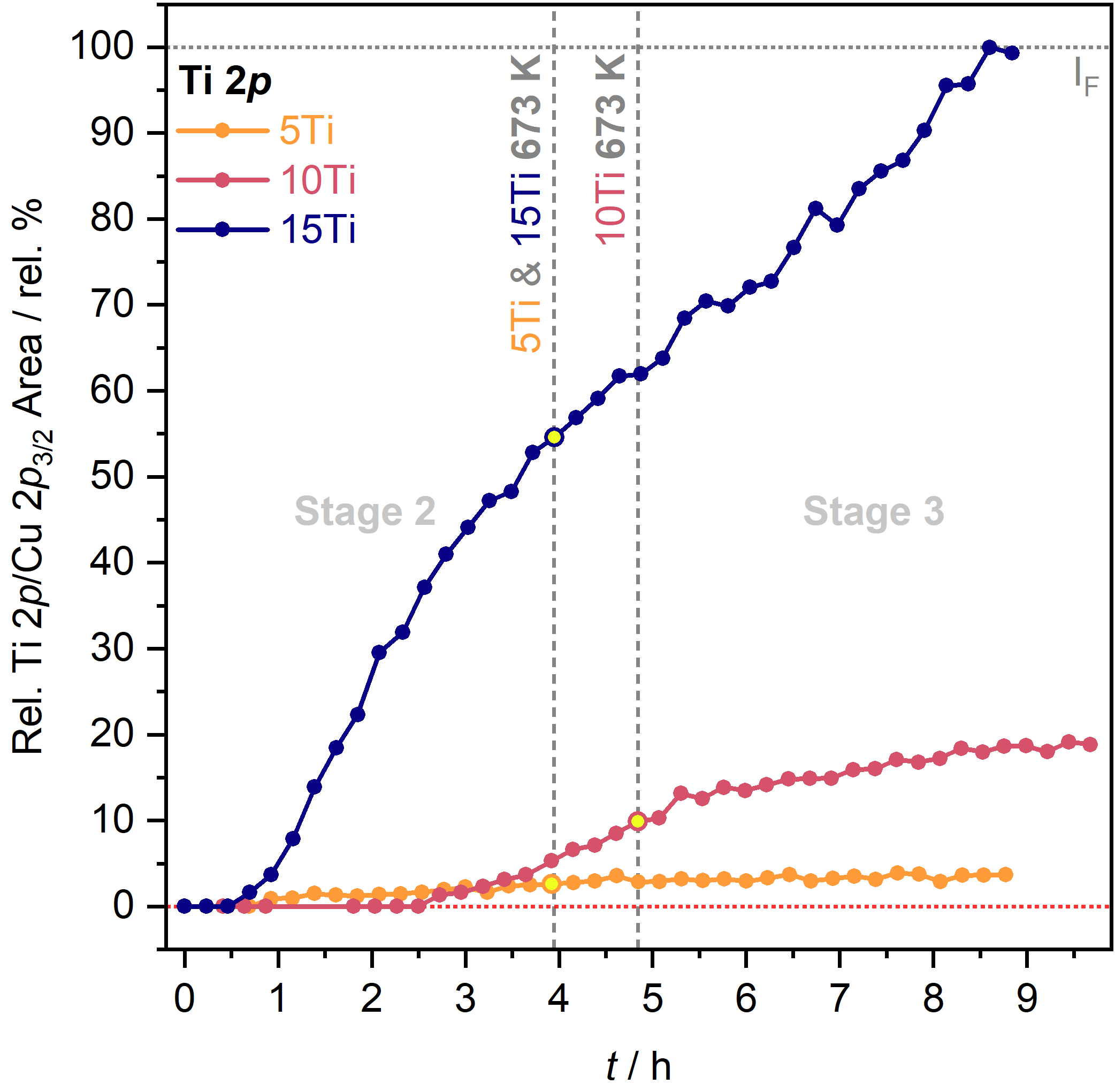}
    \caption{Relative Ti concentration profile as a function of time, \textit{t} collected across the measurement window for all three samples, determined from peak fitting the Ti~2\textit{p} core level spectra. The yellow-filled marker for each dataset refers to the time when the 673~K holding period commences. Vertical guidelines are also in place to mark this point for each sample. The measured Ti~2\textit{p} signal intensity for each sample is first normalised relative to the area of the Cu~2\textit{p}\textsubscript{3/2} core level measured during the same spectral cycle and then afterwards the resultant Ti~2\textit{p}/Cu~2\textit{p}\textsubscript{3/2} area is normalised relative to the final intensity of sample 15Ti (I\textsubscript{F}).}
    \label{fig:Ti2p_Quant}
\end{figure}



\cleardoublepage




\section{Ti~2\textit{p}/1\textit{s} comparison}

\begin{figure}[ht!]
\centering
    \includegraphics[keepaspectratio, width=0.8\linewidth]{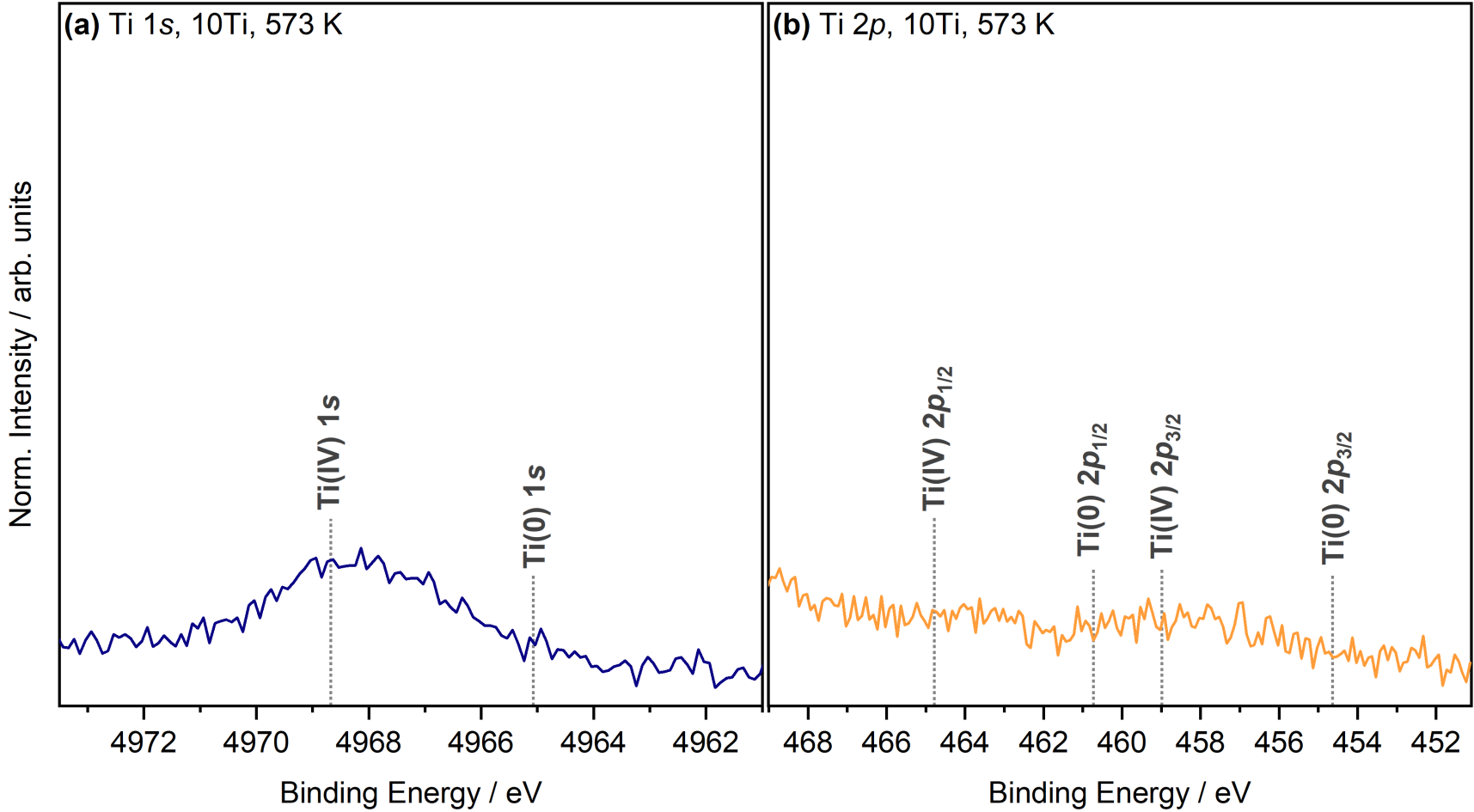}
    \caption{A comparison of the (a) Ti~1\textit{s}, and (b) Ti~2\textit{p} core level spectra recorded at 573~K (\textit{t} = 2~h) for sample 10Ti. Spectra are normalised to the signal-to-noise ratio. Guidelines are marked for the positions of the expected peaks. It is clear that the Ti~1\textit{s} is more sensitive to smaller concentrations of titanium than the Ti~2\textit{p}. Additionally, the nature of the secondary background for the Ti~2\textit{p} region means that quantification of this area is incredibly difficult and cannot be done reliably, whereas a standard XPS background can easily be applied to the Ti~1\textit{s} region.}
    \label{fig:Ti1s_2p}
\end{figure}

\cleardoublepage

\section{Depth Profile Survey Spectra}
\begin{figure}[ht!]
\centering
    \includegraphics[keepaspectratio, width=0.6\linewidth]{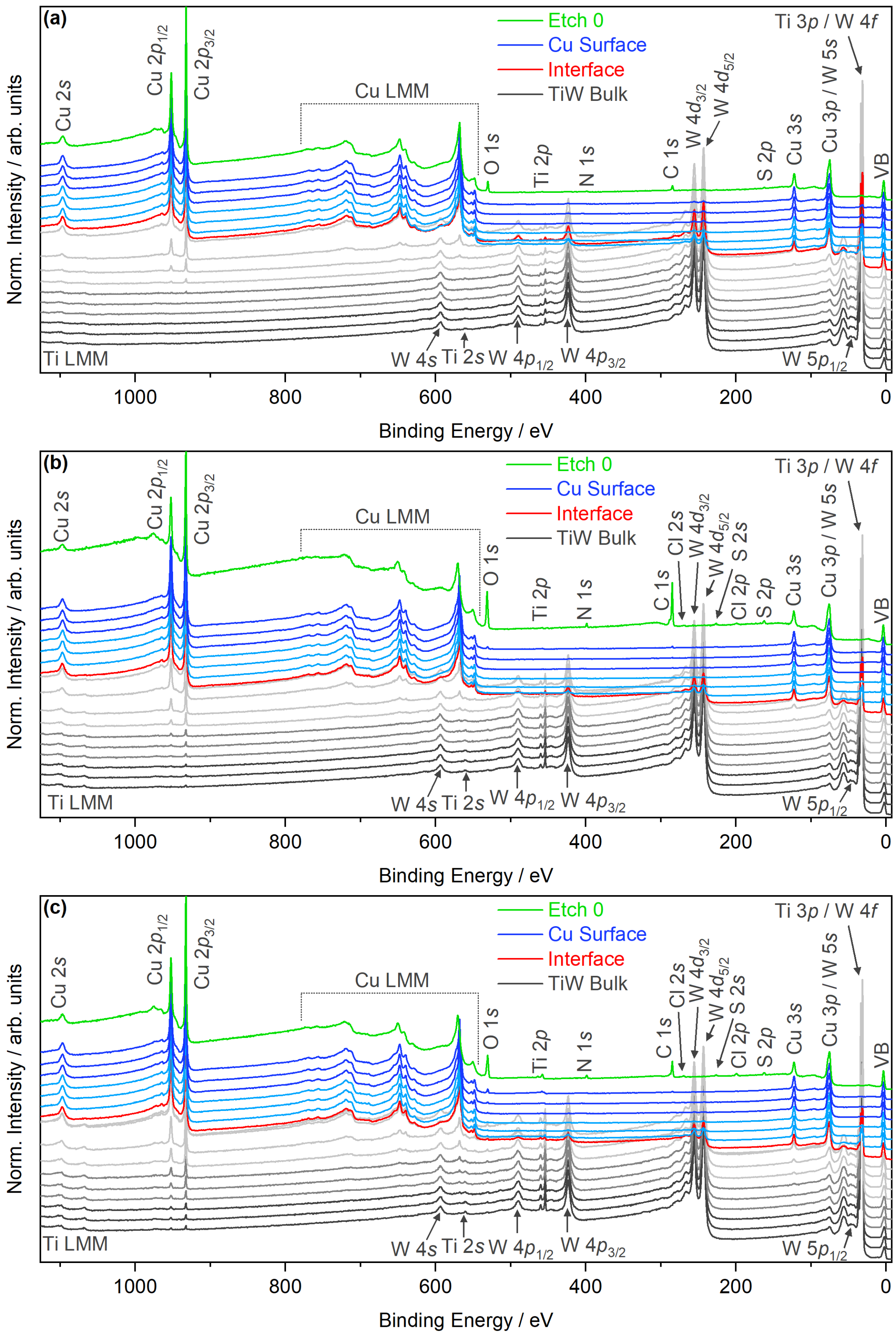}
    \caption{Survey spectra collected after each etch cycle during the post-mortem depth profile measurements for (a) 5Ti, (b) 10Ti, and (c) 15Ti samples. The top spectrum displayed in each sub-figure is taken on the as-received sample (i.e. no etch) and then the spectra collected after each cycle are stacked vertically below (going from blue to grey to black). Spectra coloured in blue are Cu-rich, black are W-rich and red is termed the ``interface'' as it marks the point where the Cu and W signals cross over in the depth profiles.}
    \label{fig:DP_surveys}
\end{figure}

\cleardoublepage

\section{References}
\bibliography{si_references}
\bibliographystyle{apsrev4-1.bst}